\documentclass[journal]{IEEEtran}
\IEEEoverridecommandlockouts

\usepackage{cite}
\usepackage{amsmath,amssymb,amsfonts}
\usepackage{mathtools}
\usepackage{amsthm}
\usepackage{algorithm}
\usepackage{algpseudocode} 
\usepackage{etoolbox}
\usepackage{graphicx,color}
\usepackage{textcomp}
\usepackage{xcolor}
\usepackage{float}
\usepackage{tikz}
\usetikzlibrary{patterns}
\usepackage{siunitx}
\usepackage{tabularx}
\usepackage{colortbl}
\usepackage{multirow}
\usepackage{array}

\usepackage{subcaption}
\usepackage{booktabs}
\usepackage{hyperref} 
\usepackage[capitalize,noabbrev]{cleveref}
\usepackage[textsize=tiny]{todonotes}
\usepackage{acronym}
\usepackage[most]{tcolorbox}
\usepackage{pdfpages}
\usepackage{setspace}
\usepackage{enumitem}

\theoremstyle{plain}
\newtheorem{theorem}{Theorem}

\newtheorem{lemma}[theorem]{Lemma}

\theoremstyle{definition}
\newtheorem{definition}[theorem]{Definition}

\theoremstyle{remark}

\definecolor{nyuviolet}{RGB}{87, 6, 140}

\newcommand{\subsf}{\sf \scriptscriptstyle}

\def\BibTeX{{\rm B\kern-.05em{\sc i\kern-.025em b}\kern-.08em
    T\kern-.1667em\lower.7ex\hbox{E}\kern-.125emX}}

\definecolor{revhdr}{RGB}{33, 85, 150}      
\definecolor{revbg}{RGB}{235, 243, 252}     
\definecolor{revtext}{RGB}{0, 0, 80}        
\definecolor{resphdr}{RGB}{0, 110, 85}      
\definecolor{respbg}{RGB}{235, 248, 243}    
\definecolor{resptext}{RGB}{0, 60, 40}      

\newtcolorbox{commentbox}[2][]{enhanced,
                            breakable,
                            colback=revbg,
                            colframe=revhdr,
                            coltitle=white,
                            coltext=revtext,
                            title={#2},
                            fonttitle=\bfseries,
                            boxrule=0.8pt,
                            arc=2pt,
                            left=4pt, right=4pt, top=2pt, bottom=2pt,
                            #1}
\newtcolorbox{responsebox}{enhanced,
                          breakable,
                          colback=respbg,
                          colframe=resphdr,
                          coltitle=white,
                          coltext=resptext,
                          title=Author Response,
                          fonttitle=\bfseries,
                          boxrule=0.8pt,
                          arc=2pt,
                          left=4pt, right=4pt, top=2pt, bottom=2pt,
                          }
\newtcolorbox{generalbox}[2][]{enhanced,
                            breakable,
                            colback=respbg,
                            colframe=resphdr,
                            coltitle=white,
                            coltext=resptext,
                            title={#2},
                            fonttitle=\bfseries,
                            boxrule=0.8pt,
                            arc=2pt,
                            left=4pt, right=4pt, top=2pt, bottom=2pt,
                            #1}
\input{defs/macros}
\newacro{ACDD}{Alamouti with cyclic delay diversity}
\newacro{URLLC}{ultra-reliable low-latency communications}
\newacro{3GPP}{third generation partnership project}
\newacro{PHY}{physical layer}
\newacro{MIMO}{multiple-input multiple-output}
\newacro{SIMO}{single-input multiple-output}
\newacro{MISO}{multiple-input single-output}
\newacro{SISO}{single-input single-output}
\newacro{MRC}{maximum-ratio combining}
\newacro{SNR}{signal-to-noise ratio}
\newacro{CP}{cyclic prefix}
\newacro{CDD}{cyclic delay diversity}
\newacro{FSC}{frequency-selective channel}
\newacro{STC}{space-time coding}
\newacro{FFT}{fast Fourier transform}
\newacro{LMMSE}{linear minimum mean-squared error}
\newacro{CFAC}{cross frequency AoA consistency}
\newacro{FER}{frame error rate}
\newacro{OFDM}{orthogonal frequency division multiplexing}
\newacro{OCDM}{orthogonal chirp division multiplexing}
\newacro{RMS}{root mean square}
\newacro{DS}{delay spread}
\newacro{FSC}{frequency-selective channel}
\newacro{CSI}{channel state information}
\newacro{LMMSE-PIC}{linear minimum mean squared error with parallel interference cancellation}
\newacro{PFE}{perfect-feedback equalizer}
\newacro{FD}{full-duplex}
\newacro{PDP}{power delay profile}
\newacro{PDF}{probability density function}
\newacro{DFT}{discrete Fourier transform}
\newacro{SDFT}{sparse DFT}
\newacro{ICI}{inter-carrier interference}
\newacro{OTFS}{orthogonal time frequency space}
\newacro{AWGN}{additive white Gaussian noise}
\newacro{SWH}{sparse Walsh-Hadamard}
\newacro{LLR}{log-likelihood ratio}
\newacro{PMF}{probability mass function}
\newacro{CRC}{cyclic redundancy check}
\newacro{PAM}{pulse amplitude modulation}
\newacro{QAM}{quadrature amplitude modulation}
\newacro{FWHT}{fast Walsh-Hadamard transform}
\newacro{MAP}{maximum a-posteriori}
\newacro{SC}{specular component}
\newacro{CFO}{carrier frequency offset}
\newacro{ISI}{inter-symbol interference}
\newacro{ZP}{zero-padding}
\newacro{EVD}{eigenvalue decomposition}
\newacro{BCJR}{Bahl, Cocke, Jelinek, and Raviv}
\newacro{WHT}{Walsh-Hadamard transform}
\newacro{APP}{a-posteriori probability}
\newacro{SILE-EPIC}{self-iterated linear equalizer with expectation propagation}
\newacro{EP}{expectation propagation}
\newacro{i.i.d.}{independent and identically distributed}
\newacro{CWCU}{component wise conditionally unbiased}
\newacro{MSE}{mean squared error}
\newacro{EXIT}{extrinsic information transfer}
\newacro{MI}{mutual information}
\newacro{PAPR}{peak-to-average power ratio}
\newacro{DFT-s}{discrete Fourier transform-spread}
\newacro{AMP}{approximate message passing}
\newacro{GAMP}{generalized \ac{AMP}}
\newacro{VAMP}{vector \ac{AMP}}
\newacro{RSC}{recursive systematic convolutional}
\newacro{QPSK}{quadrature phase-shift keying}
\newacro{CFAR}{constant false alarm rate}
\newacro{PD}{probability of detection}
\newacro{PFA}{probability of false alarm}
\newacro{RV}{random variable}
\newacro{CDF}{cumulative distribution function}
\newacro{HD-ZP}{half-duplex ZP}
\newacro{FD-CP}{full-duplex ZP}
\newacro{DFRC}{dual-function radar communication}
\newacro{SINR}{signal-to-interference noise ratio}
\newacro{ISAC}{integrated sensing and communication}
\newacro{SI}{self-interference}
\newacro{RSI}{residual self-interference}
\newacro{ADC}{analog-to-digital converter}
\newacro{DAC}{digital-to-analog converter}
\newacro{ED}{energy-detection}
\newacro{IDFT}{inverse discrete Fourier Transform}
\newacro{SFFT}{symplectic finite Fourier transform }
\newacro{CRB}{Cram{\'{e}}r-Rao bound}
\newacro{ZC}{Zadoff-Chu}
\newacro{RMSE}{root mean square error}
\newacro{UW}{unique word}
\newacro{GFDM}{generalized frequency division multiplexing}
\newacro{RRC}{root-raised cosine}
\newacro{UB}{upper bound}
\newacro{CEF}{channel estimation field}
\newacro{TRX}{transceiver}
\newacro{IF}{intermediate frequency}
\newacro{RF}{radio frequency}
\newacro{FPGA}{field programmable gate arrays}
\newacro{SDR}{software-defined radio}
\newacro{UWB}{ultra wideband}
\newacro{FR3}{frequency range 3}
\newacro{PCB}{printed circuit board}
\newacro{SMA}{SubMiniature version A}
\newacro{MUSIC}{multiple signal classification}
\newacro{CIR}{channel impulse response}
\newacro{FR}{Frequency Range}
\newacro{mmWave}{millimeter wave}
\newacro{LoS}{line-of-sight}
\newacro{AoD}{angle-of-departure}
\newacro{ESNR}{estimation SNR}
\newacro{AoA}{angle-of-arrival}
\newacro{SDNR}{signal-to-DMC-noise ratio}
\newacro{ULA}{uniform linear array}
\newacro{DMC}{dense multipath component}
\newacro{ML}{maximum-likelihood}
\newacro{IFFT}{inverse fast Fourier transform}
\newacro{LM}{Levenberg-Marquardt}
\newacro{ACF}{autocorrelation function}
\newacro{UWB}{ultra-wideband}
\newacro{SLAM}{simultaneous localization and mapping}
\newacro{STO}{sampling time offset}
\newacro{GLRT}{Generalized Likelihood Ratio Test}
\newacro{FD-ED}{frequency-domain energy detectors}
\newacro{IOU}{intersection over union}
\newacro{FLOP}{floating point operation}
\newacro{DL}{Deep Learning}
\newacro{FAP}{False Alarm Probability}
\newacro{CNN}{Convolutional Neural Network}
\newacro{WT}{Wavelet Transform}
\newacro{PSD}{Power Spectral Density}
\newacro{MRA}{Multiresolution Analysis}
\newacro{CWT}{Continuous Wavelet Transform}

\begin{document}

\title{Computationally Efficient Signal Detection with Unknown Bandwidths}



\author{
Ali Rasteh, Sundeep Rangan


\thanks{A. Rasteh and S. Rangan are with New York University,
Brooklyn, BY (e-mail: {\tt ar7655,srangan@nyu.edu}).
Their work is supported in part by NSF grants
1952180, 2133662, 2236097, 2148293, and 1925079,
 the DARPA Prowess program, the NTIA and  industrial affilates,
of NYU Wireless.}

}

\maketitle

\begin{abstract}
Signal detection in environments with unknown signal bandwidth and time intervals is a fundamental problem in adversarial and spectrum-sharing scenarios. This paper addresses the problem of detecting signals occupying unknown degrees of freedom from non-coherent power measurements, where the signal is constrained to an interval in one dimension or a hyper-cube in multiple dimensions. A \ac{GLRT} is derived, resulting in a straightforward metric involving normalized average signal energy for each candidate signal set. We present bounds on false alarm and missed detection probabilities, demonstrating their dependence on \acp{SNR} and signal set sizes. To overcome the inherent computational complexity of exhaustive searches, we propose a computationally efficient binary search method, reducing the complexity from $O(N^2)$ to $O(N)$ for one-dimensional cases. Simulations indicate that the method maintains performance near exhaustive searches and achieves asymptotic consistency, with interval-of-overlap converging to one under constant \ac{SNR} as measurement size increases. The simulation studies also demonstrate superior performance and reduced complexity compared to contemporary neural network-based approaches, specifically outperforming custom-trained U-Net models in spectrum detection tasks.
\end{abstract}


\acresetall
\section{Introduction} \label{sec:introduction}

\IEEEPARstart{S}{ignal} detection is a fundamental problem in various fields, such as communications, radar, and biomedical engineering \cite{ali2016advances, yucek2009survey, nasser2021spectrum, janu2022machine}.  In many problems, the time interval and bandwidth that the signal occupies are not known \emph{a priori}. In these cases, signal detection must be performed in parallel with
time and bandwidth estimation.  This situation arises most obviously in adversarial scenarios where
an interfering signal can have an arbitrary center frequency, bandwidth, and time interval.
The situation may also arise in future spectrum co-existence scenarios where multiple
services use arbitrary bandwidths or frequency hopping for flexibility~\cite{yucek2009survey,lee2020detection,lin2022multi,gkelias2022gan}.
Signal detection is also potentially important in emerging systems in the upper mid-band
for co-existence between disparate systems \cite{kang2024cellular,kang2024terrestrial}.

In this paper, we consider the general problem of detecting
a signal from a set of non-coherent power measurements. The signal, when present, occupies an unknown set $S$ of the degrees of freedom.
Classical \ac{FD-ED} methods typically assume a fixed sensing bandwidth or rely on known band structures (e.g., WiFi channels, LTE sub-bands). In contrast, our method explicitly addresses the case where the signal may occupy an arbitrary and unknown interval in the frequency (or, more generally, multi-dimensional) domain, under which \ac{FD-ED} proves inadequate.
One power measurement is made on each degree of freedom,
and the set of power measurements is modeled as exponential random variables with a higher mean on the degrees of freedom in $S$. While the exponential distribution of single-bin energies under \ac{AWGN} is well established, the novelty of this work lies in addressing the unknown signal structure and developing the associated search and detection methods. We consider both one-dimensional and multi-dimensional search problems, where the signal set $S$ consists of an unknown interval or hypercube within the measurement space. The dimensions of the measurement space could be frequency, time, angle of arrival, etc.
The primary focus of this work is on non-cooperative networks, where users independently sense the spectrum using the proposed method. Nevertheless, the approach is flexible and can also be applied in cooperative scenarios where users share spectrum sensing information. The proposed method is applicable to both single-antenna and multi-antenna systems.
Since the core objective is to address the detection problem—determining the presence or absence of signals at the initial stage of the processing pipeline—a single antenna is sufficient to implement our method. However, employing multiple antennas can further improve the SNR and enhance detection performance. Importantly, the system model does not impose any coherence constraints; however, an essential calibration phase is requisite for the accurate estimation of the noise power.

Existing spectrum sensing research spans a wide range of techniques, from classical energy detection -- effective when little signal information is available -- to more sophisticated approaches such as matched filtering, feature-based detection, eigenvalue methods, and numerous machine learning and deep learning–based classifiers. Although advanced detectors can offer strong performance, they often depend on prior knowledge of signal structure or incur significant computational costs, making them less suitable for real-time, resource-constrained environments. Frequency-domain and transform-based methods further improve detection but typically assume a known band structure or operate under wideband, high-rate sampling constraints, while sub-Nyquist strategies introduce their own complexities. Overall, while many algorithms exist, they either rely on restrictive assumptions, demand substantial computation, or fail to address practical constraints. These limitations motivate our work, which aims to provide an efficient, low-complexity detection method that requires minimal prior signal information and remains effective even when bandwidth and carrier frequency are unknown.

Our contributions are as follows.
\begin{itemize}
    \item \emph{Derivation of the \ac{GLRT}}:  We derive a simple \ac{GLRT} 
    \cite{soltanmohammadi2013spectrum,li2018kernelized} for the detection problem.
    For each possible candidate signal set $S$ that the signal
    can occupy, it is shown that the \ac{GLRT} can be computed from a simple metric of the average signal energy on the interval $S$, normalized by the number of degrees of freedom in $S$.
    The \ac{GLRT} then maximizes the metric over the sets.

    \item \emph{Receiver Operating Characteristic}:   Bounds are provided
    for both the false alarm and missed detection probabilities, with the missed detection probability depending on the size of the signal set and the \ac{SNR}.

    \item \emph{Computationally efficient binary search}:  A key challenge in the 
    \ac{GLRT} is that all signal sets $S$  must be searched.  For a one-dimensional problem
    with $N$ degrees of freedom, the search over all intervals will have a complexity of $O(N^2)$.
    We present a binary search method with a complexity of $O(N)$. This efficiency is particularly critical in wideband scenarios where \ac{FD-ED} would either miss signals or require prohibitively high sampling overhead.
    
    \item \emph{Asymptotic consistency of binary search}:  Our main theoretical result (Theorem~\ref{thm:iou}) shows that under constant \ac{SNR} per degree of freedom, the interval-of-overlap goes to one almost surely as the block size $N$ approaches infinity.  
    
    \item \emph{Simulation performance}:  We conduct a number of simulations and
    show that the computationally efficient binary search method performs
    similarly to the exhaustive search.  In addition, the binary search is asymptotically 
    consistent as the interval length grows.

    \item \emph{Comparison with neural network approaches}:  Given the success of neural networks
    in image segmentation, considerable work has focused on their application in spectral detection~\cite{prasad2020downscaled,tao2021radio,vagollari2021joint,li2022new,benazzouza2024novel,zhang2024prompting}. We show that the binary search method outperforms custom-trained U-Net neural networks with considerably less complexity.
\end{itemize}

\subsection*{Related Works} \label{sec:related_works}

A comprehensive survey of various spectrum sensing methodologies can be found in \cite{ali2016advances, yucek2009survey, nasser2021spectrum, janu2022machine}. In general, existing approaches can be categorized based on the prior information available regarding the signal and the channel.

The methods proposed in this paper belong to the category of energy detection, first formalized in \cite{urkowitz2005energy}, and later refined in \cite{margoosian2015accurate, sobron2015energy, chatziantoniou2015threshold, umar2013unveiling}. These techniques are most applicable when limited or no information about the signal structure is available, relying primarily on the energy variance between signal-plus-noise and noise-only hypotheses. Conversely, when additional information is available, more specialized techniques can be employed \cite{ali2016advances}. For instance, matched filtering \cite{ma2012matched, zhang2014matched} offers optimal detection performance but strictly assumes that the transmitted signal waveform is known \textit{a priori}, making it unsuitable for non-cooperative or adversarial scenarios. Similarly, feature detection methods \cite{chaudhari2009autocorrelation, huang2013cyclostationarity, rebeiz2013optimizing} and eigenvalue techniques \cite{zeng2009eigenvalue} exploit specific signal properties, such as cyclostationarity or covariance structure, which may not be present or known in arbitrary interfering signals.

In recent years, machine learning and deep learning methods have been widely proposed for the broader problem of signal identification and spectral segmentation. While these data-driven approaches often achieve strong performance, they typically incur high computational complexity \cite{prasad2020downscaled, tao2021radio, vagollari2021joint}. This presents a significant challenge in real-time and resource-constrained applications—such as spectrum sensing for cognitive radio, adversarial communications, or radar—where large-scale data must be processed with minimal latency. Excessive algorithmic complexity in these scenarios can lead to unacceptable delays, increased energy consumption, or impractical hardware requirements.

Regarding frequency domain processing, many detectors operate on transformed data using \acp{FFT} \cite{quan2008optimal}, filter banks \cite{farhang2008filter, lin2010progressive}, or wavelets \cite{el2013improved, jindal2014wavelet, martone2002multiresolution}. Our work shares similarities with \cite{quan2008optimal} in its use of frequency-domain data; however, \cite{quan2008optimal} assumes a band structure known \textit{a priori} (e.g., specific WiFi channels). In contrast, our method explicitly addresses the more general and challenging case where the signal may occupy an arbitrary and unknown interval in the frequency domain.
Similarly, \cite{mohammadi2024parallel,kumar2022intelligent,vluadeanu2024average} presents further examples of deploying transformed data while imposing specific assumptions on signal structure.

Furthermore, edge detection techniques based on the \ac{WT} have been explored for wideband sensing \cite{el2013improved, jindal2014wavelet}. As established in~\cite{tian2006wavelet}, these methods typically treat the \ac{PSD} as a train of subbands and utilize \ac{MRA} to identify singularities (edges) between occupied and vacant bands, often relying on Lipschitz regularity to distinguish signal edges from noise~\cite{mallat2002singularity}. However, these approaches differ fundamentally from our proposed method in two key aspects: complexity and structural assumptions. First, while robust wavelet sensing often requires the \ac{CWT} or computations across multiple scales $M$, scaling as $O(M \cdot N \log N)$~\cite{stephane1999wavelet, wang2023comparisons, ieng2018complexity}, our proposed \ac{GLRT}-based binary search achieves a linear time complexity of $O(N)$. Second, wavelet methods typically assume the \ac{PSD} is piecewise smooth to ensure valid edge detection~\cite{kumar2016improved, kobayashimultiband}, an assumption that may fail under frequency-selective fading or rough signal profiles. Our method avoids these restrictive assumptions, operating robustly on non-coherent power measurements.

It should also be recognized that many transformation-based methods operate under wideband constraints, meaning that the
detection data is sampled at a higher rate than the true signal. Several sub-Nyquist strategies have been explored via compressive sensing \cite{tian2007compressed, tropp2007signal, dai2009subspace} and multi-coset sensing \cite{venkataramani2000perfect}. Unlike these approaches, our work does not attempt super-resolution of frequency or time; instead, it focuses on efficient detection within the available degrees of freedom.

Cabric et al. \cite{cabric2006spectrum,cabric2007cognitive} experimentally evaluated spectrum sensing methods, establishing scaling laws for sensing time in \ac{AWGN} channels. A key result is that the number of samples required for non-coherent detection methods grows as $1/\mathrm{SNR}^2$ \cite{urkowitz2005energy, cabric2006spectrum, cabric2007cognitive}, compared to $1/\mathrm{SNR}$ for coherent matched filtering. While we do not derive new scaling laws, our results demonstrate consistent estimation with a constant \ac{SNR} per bin as the signal length grows.

Finally, while numerous studies have addressed energy efficiency in cognitive radio, the majority focus on minimizing communication-related energy consumption rather than optimizing the computational energy of the detection algorithm itself \cite{zheng2016energy, wu2022energy, ejaz2015energy, deng2011energy, wang2011energy}. Works that have considered detection complexity have mostly focused on optimizing deep learning architectures \cite{ding2022deep,mei2023deep} or compressive sensing \cite{alam2013computationally}. In this paper, we bridge this gap by proposing a method that is algorithmically efficient by design. In our simulations, we compare it to a custom deep learning method and demonstrate that, even with significantly lower computational complexity, the proposed method offers superior detection performance.

\section{Problem Formulation}
\label{sec:problem_formulation}

\subsection{Linear Search Problem} \label{sec:linear_search}
For simplicity, we first consider a linear search problem.We are given $N$ frequency bins indexed $n=0,\ldots,N-1$ and a vector of power measurements $\bs{X}= (X_0,\ldots,X_{N-1})$.
A signal occupies an unknown interval $S=[a,b)$ where $0 \leq a \leq b \leq N$ -- see Fig.~\ref{fig:search_region}(a).
Note that $a$ may equal $b$, so that $S$ is empty.  
Given the interval $S$, the received power measurements are modeled as exponential and independent, with an expectation
\begin{equation} \label{eq:xscalar}
    \Exp[ X_n ] = \begin{cases}
        1+\gamma & \mbox{if } n \in S \\
        1 & \mbox{if } n \not\in S,
    \end{cases}
\end{equation}
where $\gamma$ represents an \ac{SNR}.  
Here, we have normalized the power values so that $\Exp(X_n) = 1$ corresponds to the
case of a noise only signal. 

\begin{figure}
    \centering

    \begin{tikzpicture}[font=\footnotesize]
        \draw[->] (0,0) -- (6,0) node[right] {};
        \draw[->] (0,0) -- (0,4) node[above] {\normalsize $\Exp(X_n)$};
        
        \draw (0,0) -- (0,-0.2) node[below] {\normalsize $0$};
        \draw (2,0) -- (2,-0.2) node[below] {\normalsize $a$};
        \draw (4,0) -- (4,-0.2) node[below] {\normalsize $b$};
        \draw (5,0) -- (5,-0.2) node[below] {\normalsize $N$};
        \draw [dashed, thick] (2,0.8) -- (2,0);
        \draw [dashed, thick] (4,0.8) -- (4,0);
        
        \draw[dashed, thick] (0.5,1) -- (-0.5,1) node[left] {\normalsize $1$};
        \draw[dashed, thick] (1.8,3) -- (-0.5,3) node[left] {\normalsize $1+\gamma$};
        
        \draw[ultra thick,blue] (0,1) -- (2,1) -- (2,3) -- (4,3) -- (4,1) 
        -- (5,1);
        \node at (3,-0.8) {\normalsize Frequency bin $n$};
    \end{tikzpicture}

    \vspace{5 pt}
    
    \scalebox{0.75}{
    \begin{tikzpicture}[font=\footnotesize]
        \draw[fill=white, draw=black, thick] (0,0) rectangle (8,8);
        
        
        \node[rotate=90, anchor=center] at (-0.9,3.7) {\large Time Sample};
        \node[anchor=center] at (4.5,-1.1) {\Large Frequency bin};
        \draw (0.02,0) -- (0.02,-0.2) node[below] {\large $0$};
        \draw (7.98,0) -- (7.98,-0.2) node[below] {\large $N$};
        \draw (3.0,0) -- (3.0,-0.2) node[below] {\large $f_a$};
        \draw (6.0,0) -- (6.0,-0.2) node[below] {\large $f_b$};
        \draw (0.0,2.5) -- (-0.2,2.5) node[left] {\large $t_a$};
        \draw (0.0,5.0) -- (-0.2,5.0) node[left] {\large $t_b$};
        \draw [dashed, thick] (3.0,2.5) -- (3.0,0);
        \draw [dashed, thick] (6.0,2.5) -- (6.0,0);
        \draw [dashed, thick] (3.0,2.5) -- (0.0,2.5);
        \draw [dashed, thick] (3.0,5.0) -- (0.0,5.0);
        
        \filldraw[color=cyan!20, draw=black, thick] (3.0,2.5) rectangle (6.0,5);
        \node[anchor=center] at (4.5,3.8) {\large $\Exp(X_n)=1+\gamma$};
        \node[anchor=center] at (4.5,6.5) {\large $\Exp(X_n)=1$};
        
    \end{tikzpicture}
    }
    

    

    
    \caption{Detection problem examples:  
    Top: $d=1$ example for 
    finding an interval $[a,b)$ of signal energy
    in $N$ frequency bins;
    Bottom:  $d=2$ example of finding
    a bounding box in time-frequency.}
    \label{fig:search_region}
\end{figure}

The assumption of the exponential distribution 
on the energy per bin is a simplification that can arise as follows:
Suppose $X_n = |H_nU_n + W_n|^2$
, where $H_n$ is a complex channel gain, $U_n$ is a complex transmitted symbol, and $W_n$ is complex Gaussian noise.
Since the receiver is not synchronized with the transmitter, the complex symbol
$U_n$ on any particular frequency bin
can be modeled as complex Gaussian noise.
If we treat the channel gain as deterministic $H_n$, then $X_n$ will be complex Gaussian with variance  $|H_n|^2\Exp|U_n|^2 + \Exp|W_n|^2$. In particular, \eqref{eq:xscalar} holds if the noise is normalized to unity, such that $\Exp|W_n|^2=1$ and $\gamma=|H_n|^2\Exp|U_n|^2$ are the received energy in bins $n \in S$.

Note that there are two implicit critical assumptions:
First, since $\gamma$ is constant over the
entire signal region, we are implicitly assuming
 that the signal has a constant PSD and that the channel is flat.  An interesting
avenue for future work is to investigate what happens in cases of frequency-selective fading.
Second, by assuming that the noise is normalized to unity, we have assumed that the noise level is known; hence, we assume that the system has undergone a noise estimation process.

For exposition, we have considered the  single antenna case.  The distribution would need to be modified for the multi-antenna case.  Specifically, the total noise across $N$ antennas will be a sum of exponential random variables, which follows a chi-squared distribution.

Our problem is twofold: First,
we wish to determine whether a signal is present.
We model this first problem as a hypothesis testing
problem.  We denote the null hypothesis by the event $H=0$, which corresponds to the case when there is no signal. That is, for all $n$, $X_n$ are i.i.d. 
exponentially distributed 
\begin{equation}
    \Exp[X_n] = 1 \quad \mbox{for all } n.
\end{equation}
The case of the null hypothesis ($H=0$) is identical to
\eqref{eq:xscalar} in cases where the \ac{SNR} is
$\gamma  = 0$.  We will denote by $H=1$
the event that there is a signal in some
signal region $S$ with a \ac{SNR} $\gamma$.

Second, if a signal is detected (that is, it is
estimated that $H=1$), we wish
to estimate the \ac{SNR} $\gamma$ and the signal interval $S$.

\subsection{Multi-Dimension Extension} \label{sec:multi_dimension}
The problem is naturally extended to dimensions $d \geq 1$.  In this case,
$\bs{X}$ is a tensor of $d$ -th order with components $X_{n_1,\ldots,n_d}$ with $n_i \in [0,N_i)$.
For example, $\bs{X}$ could be measurements
over $N_0$ time bins and $N_1$ frequency bins,
as you may obtain from $N_0$ \acp{FFT}, each \ac{FFT}
of length $N_1$.
In the multi-dimensional case, 
the set $S$ is a hyper-cube:
\begin{equation} \label{eq:Smulti}
    S = \left\{  (n_1,\ldots,n_d) ~|~ a_i \leq n_i < b_i ~\forall i=1,\ldots,d 
    \right\},
\end{equation}
for some left and right intervals $\bs{a}=(a_1,\ldots,a_d)$ and $\bs{b}=(b_1,\ldots,b_d)$.
As an example, Fig.~\ref{fig:search_region}(b)
shows a $d=2$ case for a time-frequency search.
The signal region corresponds to a bounding box
in time and frequency.
Again, we assume that, given $S$, $X_{n_1,\ldots,n_d}$ are exponential and independent with:
\begin{equation} \label{eq:xvector}
    \Exp[ X_{n_1,\ldots,n_d} ]
        = \begin{cases}
        1+\gamma & \mbox{if } (n_1,\ldots,n_d) \in S \\
        1 & \mbox{if } (n_1,\ldots,n_d)  \not\in S,
    \end{cases}
\end{equation}
where, again, $\gamma$ represents an \ac{SNR}. 
To use a uniform notation between the vector and tensor cases, we will let
$n = (n_1,\ldots,n_d)$ be a $d$-dimensional tensor index and write the components
of the tensor as $X_n= X_{n_1,\ldots,n_d}$.

Again, the problem in the multi-dimensional case
is to detect whether a signal is present, and if so,
to estimate the signal region $S$
and \ac{SNR} $\gamma > 0$.


\section{Likelihood Ratio Test} \label{sec:likelihood_ratio_test}

\subsection{Likelihood Function}
We follow a standard likelihood ratio detector
for the signal detection problem.
Let $H=0,1$ denote the true signal hypothesis:
\begin{equation} \label{eq:Htrue}
    H = \begin{cases}
        0 & \mbox{no signal present;} \\
        1 & \mbox{signal is present}.
    \end{cases}
\end{equation}
We compute an estimate $\wh{H}=0,1$
of $H$ as follows:
Let $p(\bs{X}| S,\gamma)$ denote the \ac{PDF} of the measurements of $\bs{X}$ 
for a given signal region $S$ and
\ac{SNR} $\gamma$. Also, let $p_0(\bs{X}) := p(\bs{X}|\emptyset,0)$
denote the conditional \ac{PDF} for the null hypothesis
$H=0$ when the signal is not present (i.e., no \ac{SNR}).
Let $J(S,\gamma)$
denote the log likelihood ratio 
\begin{equation}
 J(S,\gamma) := \log \left[ \frac{p(\bs{X}|S,\gamma)}{p_0(\bs{X})} \right].
\end{equation}
Since the parameters $\gamma$ and $S$ are not known,
we use a modified version of the 
\ac{GLRT}
\cite{soltanmohammadi2013spectrum,li2018kernelized}:  We compute the maxima:
\begin{equation} \label{eq:Jmax}
    J^*_{\ell} := \argmax_{\substack{S,\gamma\geq 0\\ |S|=\ell}} J(S,\gamma) 
\end{equation}
where $|S|$ is the cardinality (number of elements)
of $S$.  Hence, $J^*_{\ell}$ is the maximum over all
$\gamma > 0$ and sets $S$ with $|S|=\ell$.  We
then assume a detector of the form:
\begin{equation} \label{eq:Hest}
    \wh{H}= \begin{cases}
        1 & J^*_{\ell} \geq t_{\ell} \mbox{ for some } \ell\\
        0 & J^*_{\ell} < t_{\ell} \mbox{ for all } \ell,
    \end{cases}
\end{equation}
where $t_\ell$ are a set of thresholds
that depend on the set size $S$.  
In the standard \ac{GLRT}, the threshold levels
$t_\ell$ are equal. However, as we will see below,
having a set size-dependent threshold will enable
a simpler analysis for the false alarm probability.

For each $\ell$, we can
compute $J^*_\ell$ in \eqref{eq:Jmax},
in two steps:  First, for each $S$, we maximize
the likelihood ratio $J(S,\gamma)$ over $\gamma$:
\begin{equation} \label{eq:Js}
    J(S) := \max_{\gamma \geq 0} J(S,\gamma).
\end{equation}
Then, $J^*_{\ell}$ can be computed by maximizing over
$S$:
\begin{equation} \label{eq:JmaxS}
    J^*_{\ell} = \max_{|S|=\ell} J(S).
\end{equation}
Our first lemma provides a simple expression
for the likelihood ratio and maximization
over $\gamma$:

\begin{lemma} \label{lem:likelihood} 
The likelihood is given by:
\begin{equation} \label{eq:Jsgam}
  J(S,\gamma) = |S| \left[ \wb{X}_S\left[ \frac{\gamma}{1+\gamma} \right]  - \log(1+\gamma) \right].
\end{equation}
where $\wb{X}_S$ is the average value of $X_n$ on the set $S$:  
\begin{equation} \label{eq:Xs}
 \wb{X}_S = \frac{1}{|S|} \sum_{n \in S} X_n,
\end{equation}
The likelihood 
maximized over the \ac{SNR} is given by:
\begin{equation} \label{eq:Js_max}
    J(S) = |S| \left[ 
    \wb{X}_S^+ -1- \log(\wb{X}_S^+) \right].
\end{equation}
where 
\begin{equation}
    \wb{X}_S^+ = \max\{\wb{X}_S, 1\}.
\end{equation}
\end{lemma}
\begin{proof}
    See Appendix~\ref{sec:likelihood_proof}.
\end{proof}

Using this lemma, we can rewrite the \ac{GLRT} as follows:
Let $\varphi(x)$ be the function:
\begin{equation} \label{eq:varphidef}
    \varphi(x) = x-1 - \log(x), \quad 
    x \geq 1.
\end{equation}
Lemma~\ref{lem:likelihood} shows that the \ac{GLRT}
\eqref{eq:Hest} can be written as $\wh{H}=1$
when 
\begin{equation} \label{eq:varphicond}
    \ell\varphi(\wb{X}^+_S) \geq t_{\ell} \mbox{ for some }
    |S|=\ell.
\end{equation}
Since $\varphi(x)$ is increasing for $x \geq 1$,
we can define 
\begin{equation} \label{eq:uidef}
    u_{\ell} = \varphi^{-1}(t_{\ell}/\ell),
\end{equation}
and the \ac{GLRT} estimator \eqref{eq:Hest}
is equivalent to:
\begin{equation} \label{eq:Hestu}
    \wh{H}= \begin{cases}
        1 & X^+_S \geq u_{|S|} \mbox{ for some } S\\
        0 & X^+_S < u_{|S|} \mbox{ for all } S.
    \end{cases}
\end{equation}

\subsection{False Alarm Probability} \label{sec:fa_probability}
Our next result bounds the false alarm
probability 
\begin{equation} \label{eq:Pfa}
    P_{\subsf FA} = \mathbb{P}(\wh{H}=1|H=0)
\end{equation}
as a function of the threshold levels $u_\ell$
in the \ac{GLRT} \eqref{eq:Hestu}.
This bound will allow us to select the
threshold levels.
To state the bound,  
consider
the modified \ac{GLRT} detector \eqref{eq:Hest}
(or, equivalently,
the \ac{GLRT} detector in \eqref{eq:Hestu})
over a set of hyper-cube regions in
\eqref{eq:Smulti}.  Define 
\begin{equation} \label{eq:Nprod}
    N = N_1N_2 \cdots N_d.
\end{equation}
Each candidate signal region $S$
is defined by a hyper-cube \eqref{eq:Smulti}
with boundaries $\bs{a}$ and $\bs{b}$.  Since
there are
$N$ in \eqref{eq:Nprod} choices for $\bs{a}$
and $\bs{b}$, and $\bs{b}_i \geq \bs{a}_i $, so there are at most $\frac{N^2}{2}$ possible
set hyper-cube $S$.  We can then apply a union
bound over this set of hypotheses
to bound the false alarm probability.

\begin{lemma} \label{lem:pfa}  
Consider
the modified \ac{GLRT} detector \eqref{eq:Hestu}
over a set of hyper-cube regions in
\eqref{eq:Smulti}.  Assume that all thresholds satisfy
$u_i \geq 0$.
Then, the probability of false alarm 
\eqref{eq:Pfa} is bounded above by
\begin{equation} \label{eq:pfabnd}
    P_{\subsf FA} \leq \frac{N^2}{2} \max_{\ell=1,\ldots,N}
    F\left(2\ell u_{\ell}; 2 \ell \right),
\end{equation}
where $N$ is defined in \eqref{eq:Nprod}
and $F(s,\nu)$ is the complementary \ac{CDF}
for a chi-squared random variable with $\nu$
degrees of freedom.  That is,
\begin{equation} \label{eq:ccdf_cs}
    F(s;\nu) = \mathbb{P}( Z_\nu \geq s),
\end{equation}
where $Z_\nu$ is a chi-squared random variable
with $\nu$ degrees of freedom.
\end{lemma}
\begin{proof} See Appendix~\ref{sec:pfa_proof}.
\end{proof}

\begin{figure}
    \centering
    \includegraphics[width=0.9\linewidth]{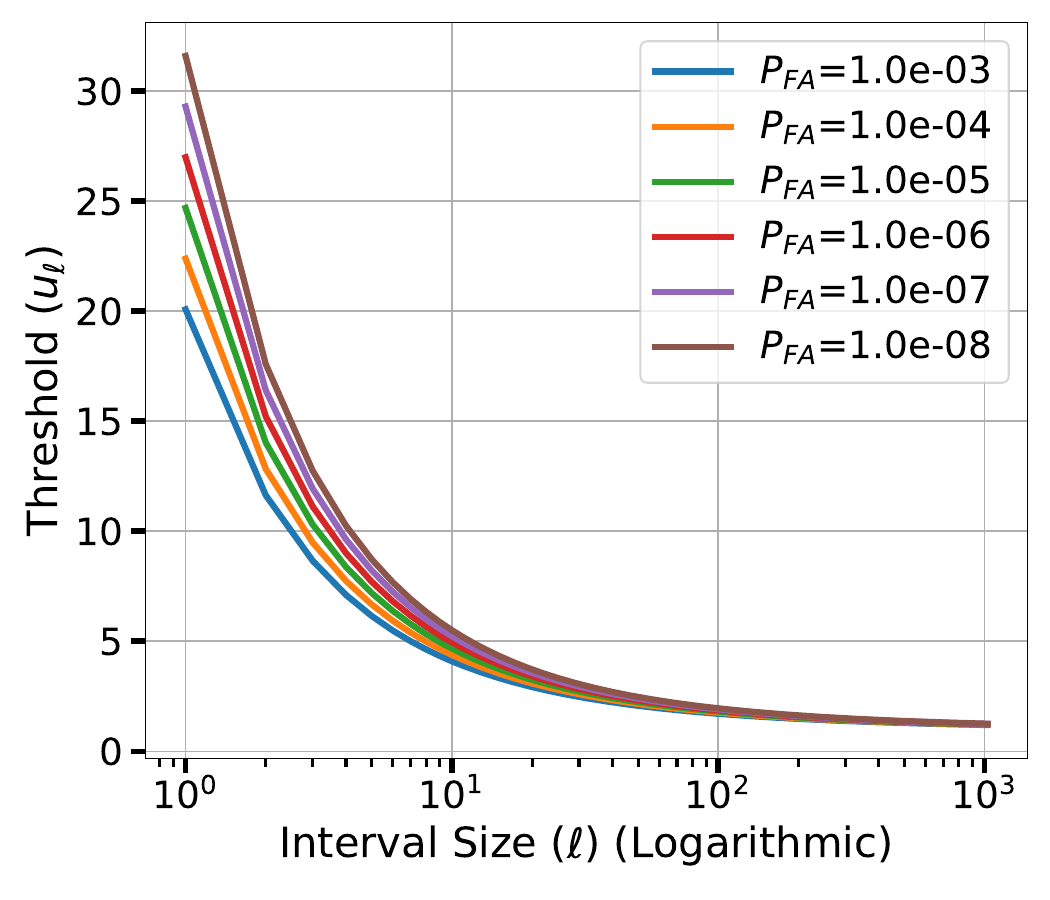}
    \caption{Threshold values $u_{\ell}$ from equation~\ref{eq:uisolve} as a function of signal interval size, showing higher thresholds for shorter intervals due to increased noise power variance.}
    \label{fig:thr_vs_dof}
\end{figure}

The lemma~\ref{lem:pfa} provides a simple recipe
for computing the threshold levels $u_{\ell}$ for the \ac{GLRT}
detector in \eqref{eq:Hestu}, or equivalently,
the threshold levels $t_{\ell}$ for the \ac{GLRT} detector in
\eqref{eq:Hest}.
First, we set a desired false alarm probability
target $P_{\subsf FA}$.  Then, each $u_{\ell}$ should satisfy
\begin{equation} \label{eq:Pfasolve1}
 \frac{2 P_{\subsf FA}}{N^2}  = 
    F\left(2 \ell u_{\ell}; 2 \ell \right).
\end{equation}
We solve \eqref{eq:Pfasolve1} by first computing
the value $u_{\ell}$ with the inverse complementary \ac{CDF}:
\begin{equation} \label{eq:uisolve}
 u_{\ell} = \frac{1}{2\ell}F^{-1}\left(\frac{2 P_{\subsf FA}}{N^2}; 2\ell\right),
\end{equation}
Then, we compute $t_{\ell}$ from \eqref{eq:uidef}
\begin{equation} \label{eq:tisolve}
    t_{\ell} = \ell\varphi\left( u_{\ell} \right).
\end{equation}
Fig.~\ref{fig:thr_vs_dof} illustrates the threshold values computed using equation~\ref{eq:uisolve} for varying signal interval sizes. For smaller intervals, the threshold on the mean power is higher, which is expected since the mean noise power exhibits greater variance over shorter intervals.

\subsection{Probability of Missed Detection} \label{sec:md_probability}
The probability of missed detection is 
\begin{align}
    \MoveEqLeft P_{\subsf MD} = \mathbb{P}(\wh{H}=0|H=1),
\end{align}
which is generally a function of the true
signal region $S$ and \ac{SNR} $\gamma$.
The following lemma provides a bound on the
missed detection probability.

\begin{lemma}  \label{lem:pmd}
If the true signal region is $S$ and true
\ac{SNR} is $\gamma$, the probability of missed detection is bounded by:
\begin{equation} \label{eq:pmdbnd}
    P_{\subsf MD} \leq 1 - F\left(\frac{2\ell u_{\ell}}{1+\gamma}; 2\ell \right),
\end{equation}
where $\ell=|S|$, where, as before $F(x;\nu)$
is the complementary \ac{CDF} \eqref{eq:ccdf_cs}
of a chi-squared random variable with $\nu$
degrees of freedom.
\end{lemma}
\begin{proof}  See Appendix~\ref{sec:proof_pmd}.
\end{proof}

It is important to note that changes in noise variance correspond to a scaling of the \ac{SNR} parameter $\gamma$, which is explicitly incorporated in the detection probability analysis (Lemma~\ref{lem:pmd}, equation~\ref{eq:pmdbnd}). In practice, an increase in noise variance reduces the effective \ac{SNR}, which, in turn, shifts the detection and missed detection probabilities. However, the false alarm probability remains fixed at the pre-specified level since the thresholds are chosen under the null distribution independently of the signal.



\subsection{Comparison to Known Interval Search} \label{sec:known_interval}
The primary challenge addressed in this paper is the uncertainty associated with the signal interval, denoted as $S$. To understand the implications of this uncertainty, we can compare the \ac{GLRT} detector with an equivalent detector that has prior knowledge of the signal interval $S$. Specifically, consider a scenario in which an oracle provides a singular true potential signal set, denoted as $S_0$.
In this context, we consider two hypotheses: the null hypothesis $H_0$, where the signal is absent, and the alternative hypothesis $H_1$, which suggests the presence of a signal within the signal set $S_0$. Define $\ell_0 = |S_0|$ as the size of the known signal set, and let $u_{0}$ represent the associated threshold value as specified in equation~\ref{eq:Hestu}.
The resulting \ac{GLRT} detector will simply be:

\begin{equation}
    \wh{H}= \begin{cases}
        1 & X^+_{S_0} \geq u_{0}\\
        0 & X^+_{S_0} < u_{0}.
    \end{cases}
\end{equation}
The variable \(u_{0}\) denotes the threshold level associated with \(S_{0}\).
According to Section~\ref{sec:fa_probability}, the expression for \(P_{\subsf FA}^0\) and \(u_{0}\) is given by:
\begin{align}
    P_{\subsf FA}^0 &= F(2\ell_0 u_{0}; 2\ell_0), \\
    u_{0} &= \frac{1}{2\ell_0}F^{-1}\left(P_{\subsf FA}^0; 2\ell_0 \right),
\end{align}
Similarly, as delineated in Section~\ref{sec:md_probability}, the expression for \(P_{\subsf MD}^0\) may be formulated as follows:
\begin{align}
    P_{\subsf MD}^0 &= 1-F(\frac{2\ell_0 u_{0}}{1+\gamma}; 2\ell_0),
\end{align}
Consider the selection of \( u_{0} \) such that the false alarm probability, \( P_{\text{FA}} \), is equal to the predefined threshold \( P_{\text{FA}}^0 \). It can be anticipated that the imposition of a more stringent constraint on the false alarm probability \( P_{\subsf FA} \) will consequently enhance performance concerning the missed detection probability \( P_{\subsf MD} \). This improvement reflects the trade-off incurred due to the introduction of interval estimation uncertainty in the proposed methodology. The experimental results concerning this section are elaborated in section~\ref{sec:known_interval_effect}

\subsection{Behavior in Multi-Signal Environments}
While the system model in Section~\ref{sec:problem_formulation} explicitly assumes a single continuous signal interval $S$, it is valuable to characterize the \ac{GLRT} detector's behavior in environments containing multiple disjoint signals. In such scenarios, the algorithm seeks the single interval $S$ that maximizes the metric defined in Eq.\ref{eq:Js_max}. Consequently, the detection outcome depends on the separation distance between the signals.
In cases where signals are widely separated by a significant span of noise, the algorithm exhibits a winner-takes-all behavior. Merging two distant signals into a single estimate would require including the intervening noise-only observations, which significantly lowers the average power $\overline{X}_{S}$. Since the term $[\overline{X}_{S} - 1 - \log(\overline{X}_{S})]$ in the likelihood function diminishes rapidly as $\overline{X}_{S}$ approaches the noise floor of 1, the penalty for including a large noise gap typically outweighs the linear gain in degrees of freedom $|S|$. As a result, the detector converges on the single signal component that yields the highest individual likelihood score. Conversely, if the signals are closely spaced, the algorithm is likely to merge them into a single aggregate interval, as the likelihood gain from capturing the energy of the second signal outweighs the penalty introduced by the short noise segment between them.

\section{Computationally Efficient Binary Search}   \label{sec:binary}

\subsection{Dyadic Interval Detection}
The \ac{GLRT} detector in \eqref{eq:Hest} requires
searching over all possible signal regions $S$.
We will call this method \emph{exhaustive \ac{ML}}.
As discussed above, there are $\approx N^2/2$
such signal regions where $N$ is given by
the product \eqref{eq:Nprod}.  Hence,
the complexity of exhaustive \ac{ML} will be $O(N^2)$, which may
be challenging, particularly at high sample rates.

In this section, we propose a simplified search
method for the linear search case where $d=1$. 
 Exhaustive \ac{ML} in this case has a complexity
$O(N^2)$.  The proposed method will have a complexity $O(N)$.

The key to the algorithm is to perform a binary search on a set of
\emph{dyadic intervals}.  We assume $N=2^M$
for some $M\geq 0$.  The dyadic intervals
are \begin{equation}
    I_{m,i} = [i2^m, (i+1)2^m) \subseteq [0,1,\ldots,N),
\end{equation}
where $m=0,\ldots,M$ and $i=0,\ldots,2^{M-m}-1$.
We will call $I_{m,i}$ the $i$-th dyadic interval at the $m$-th level. 
The proposed algorithm, which we call \emph{binary search}, operates in two phases:
\begin{itemize}
    \item \emph{Initial interval detection} 
    where we search the set of dyadic intervals to detect whether a signal is present.
    \item \emph{Binary interval estimation}:  
    If
    a signal is detected in the dyadic interval search, the most likely interval is estimated using a binary search.
\end{itemize}

\begin{algorithm}
\caption{Initial Interval Search}\label{alg:dyad_search}
\begin{algorithmic}[1]
\Require Interval length $N = 2^M$
\Require Power measurements $X_n, n=0,\ldots N-1$
\Require Threshold levels $u_k$, $k=1,2,4,\ldots,2^M$
\State $\wh{H} \gets 0$
\For{$m=0,\ldots,M$}
    \State $\ell \gets 2^m$ // interval length
    \State $k \gets 2^{M-m}$ // number of intervals
    
    \State $Z^*_m \gets 0$.
    \For{$i=0,\ldots, k-1$}
        \If{$m=0$} 
            \State $Z_{0,i} \gets X_i$
        \Else 
            \State $Z_{m,i} \gets (Z_{m-1,2i}+Z_{m-1,2i+1})/2$.
        \EndIf
        
        \If{$Z_{m,i} \geq u_k$}
            \State $Z^*_m \gets Z_{m,i}$ 
            \State $i^*_m \gets i$
        \EndIf
    \EndFor
    \If{$Z^*_{m} \geq u_k$}
        \State $\wh{H} \gets 1$
        \State $J^*_m \gets \ell( Z^*_m -1 -\log(Z^*_m))$
    \EndIf
\EndFor
\If{$\wh{H} = 1$}
    \State $\wh{m} \gets \argmax_m J^*_m$
    \State $\wh{i} \gets i^*_{\wh{m}}$
    \State \Return ($\wh{H}=1$ $\wh{m}$, $\wh{i}$)
\Else
    \State \Return $\wh{H}=0$
\EndIf
\end{algorithmic}
\end{algorithm}

In this subsection, we describe the first
stage, namely the dyadic interval search.
The precise steps are shown in Algorithm~\ref{alg:dyad_search}.
In this algorithm, we start with \( N \) intervals, each representing a power measurement bin \( X_i \). At each stage, we sequentially merge adjacent intervals and compute \( X_S \) for all resulting intervals.
It can be verified that:
\begin{equation} \label{eq:Zmi}
    Z_{m,i} = \wb{X}_S \mbox{ for } S = I_{m,i},
\end{equation}
So, the algorithm computes the mean energy
over all the dyadic intervals.  The resulting
estimate is then:
\begin{equation} \label{eq:Hest_dyad}
    \wh{H} = \begin{cases}
        1 & \mbox{if } \wb{X}_S \geq u_{|S|}
        \mbox{ for some } S=I_{m,i} \\
        0 & \mbox{if } \wb{X}_S < u_{|S|}
        \mbox{ for all } S=I_{m,i}.
    \end{cases}
\end{equation}
Hence, Algorithm.~\ref{alg:dyad_search}
produces a hypothesis estimate $\wh{H}$
that matches the estimate for the \ac{GLRT}
detector \eqref{eq:Hestu}, except that
the search is only over the dyadic intervals
$S = I_{m,i}$ as opposed to all subsets.

\subsection{Interval Estimation}

If a signal has been detected by Algorithm~\ref{alg:dyad_search},
we then obtain an estimate for the interval by Algorithm~\ref{alg:int_estimation}.
This algorithm takes as inputs the binary energy values $Z_{m,i}$
from Algorithm.~\ref{alg:dyad_search}.
If the signal length is $N=2^M$, the algorithm proceeds in $M$ iterations
indexed by $t=0,\ldots,M-1$.  In each iteration, it finds an estimate for the
interval of the form:
\begin{equation}
    \wh{S}^{(t)} = [\wh{a}^{(t)}, \wh{b}^{(t)}),
\end{equation}
where the left and right boundaries
$\wh{a}^{(t)}$ and $\wh{b}^{(t)}$ are of the form:
\begin{equation}
    \wh{a}^{(t)} = \wh{i}^{(t)}\ell_t,
    \quad \wh{b}^{(t)} = \wh{j}^{(t)}\ell_t,
    \quad 
    \ell_t = 2^{M-t},
\end{equation}
so that they are restricted to discrete values:
\[
    \wh{a}^{(t)}, \wh{b}^{(t)} \in \{0, \ell_t,
    2\ell_t, \cdots, N\}.
\]
We call the spacing
$\ell_t$ the \emph{sub-interval length}
in iteration $t$.
Hence, the algorithm finds iteratively finer
estimates of the interval boundaries since 
the sub-interval length $\ell_t$
decreases as $t$ increases.  Initially, the algorithm takes the estimates
\begin{equation}
    \wh{S}^{(0)} = [\wh{a}^{(0)}, \wh{b}^{(0)})
    =[0,N),
\end{equation}
corresponding to the entire interval.

\begin{algorithm}
\caption{Binary Interval Estimation}\label{alg:int_estimation}
\begin{algorithmic}[1]
\Require Interval length $N = 2^M$
\Require Dyadic powers $Z_{m,i}$
\State $\wh{i}^{(0)} \gets 0$  // initial left index
\State $\wh{j}^{(0)} \gets 1$  // initial right index
\State $\wh{Z}^{(0)} = Z_{M,0}$, // initial average energy
\State $\wh{L}^{(0)} \gets N=2^M$ // initial interval length
\State $\ell_0 \gets 2^M$ // initial sub-interval length
\State $Z^+=\max\{\wh{Z}^{(0)},1\}$ 
\State $J_{\rm max} = \wh{L}^{(0)}[Z^+-1-\ln(Z^+)]$ // initial objective
\State

\For{$t=0,\ldots,M-1$}

    \State // Get layer info
    \State $m \gets M-t-1$ // layer    
    \State $\ell_{t+1} \gets 2^m$ // sub-interval length

    \State
    \State // Optimization over left boundary
    \State $\tilde{Z}^{(t)} \gets \wh{Z}^{(t)}$,
    $\wh{i}^{(t+1)} \gets 2\wh{i}^{(t)}$,
    $\tilde{L}^{(t)} \gets \wh{L}^{(t)}$
    \For {$\delta \in \{-1, 1\}$}
        \State $i' \gets 2\wh{i}^{(t)} + \delta$
        \State $L' \gets \wh{L}^{(t)} - \delta \ell_{t+1}$
        \State $Z' \gets (\wh{L}^{(t)} \wh{Z}^{(t)} - \delta \ell_{t+1} Z_{m,i'})/L'$
        \State $J' \gets L'[ (Z^+-1) - \ln(Z^+) ], Z^+=\max\{Z',1\}$
        \If {$J' > J_{\rm max}$, $L' \geq 0$, \textbf{and} $i'\geq 0$}
            \State $J_{\rm max} \gets J'$
            \State $\wh{i}^{(t+1)} \gets i'$
            \State $\tilde{Z}^{(t)} \gets Z'$,  $\tilde{L}^{(t)} \gets L'$
        \EndIf
    \EndFor

    \State
    \State // Optimization over right boundary
    \State $\wh{Z}^{(t+1)} \gets \tilde{Z}^{(t)}$,
    $\wh{j}^{(t+1)} \gets 2\wh{j}^{(t)}$,
    $\wh{L}^{(t+1)} \gets \tilde{L}^{(t)}$
    \For {$\delta \in \{-1, 1\}$}
        \State $j' \gets 2\wh{j}^{(t)} + \delta$
        \State $L' \gets \tilde{L}^{(t)} + \delta \ell_{t+1}$
        \State $Z' \gets (\wh{L}^{(t)} \wh{Z}^{(t)} + \delta \ell_{t+1} Z_{m,j'})/L'$
        \State $J' \gets L'[ (Z^+-1) - \ln(Z^+) ], Z^+=\max\{Z',1\}$
        \If {$J' > J_{\rm max}$, $L'> 0$, \textbf{and} $j'\leq 2^{t+1}$}
            \State $J_{\rm max} \gets J'$
            \State $j^{(t+1)} \gets j'$
            \State $\wh{Z}^{(t+1)} \gets Z'$, $\wh{L}^{(t+1)} \gets L'$
        \EndIf
    \EndFor
\EndFor
\end{algorithmic}
\end{algorithm}

\begin{figure*}[]
  \centering

    \begin{tikzpicture}
    \definecolor{greenNode}{RGB}{0,255,0}
    \definecolor{redNode}{RGB}{255,0,0}
    \definecolor{stageColor}{RGB}{255,102,255}
    
    \foreach \y/\stage in {0/Stage 0, 1/Stage 1, 2/Stage 2, 3/Stage 3, 4/Stage 4} {
        \node[draw, thick, fill=stageColor, text=black, rounded corners, inner sep=4pt] at (-1.3, -\y) {\textbf{\stage}};
    }

    \foreach \y in {0,...,4} {
        \pgfmathtruncatemacro{\z}{2^\y}   
        \pgfmathtruncatemacro{\t}{2^(4-\y)}   
        
        \foreach \x in {0,...,\numexpr \z-1 \relax} {  
            \pgfmathtruncatemacro{\m}{\x * \t}
            \pgfmathtruncatemacro{\n}{(\x+1) * \t - 1}

            \ifnum\y=0
                \ifodd\x
                    \definecolor{boxColor}{RGB}{224,224,224} 
                \else
                    \definecolor{boxColor}{RGB}{224,224,224} 
                \fi
            \fi
            \ifnum\y=1
                \ifodd\x
                    \definecolor{boxColor}{RGB}{192,224,224} 
                \else
                    \definecolor{boxColor}{RGB}{224,192,224} 
                \fi
            \fi
            \ifnum\y=2
                \ifodd\x
                    \definecolor{boxColor}{RGB}{224,224,192} 
                \else
                    \definecolor{boxColor}{RGB}{160,224,224} 
                \fi
            \fi
            \ifnum\y=3
                \ifodd\x
                    \definecolor{boxColor}{RGB}{224,160,224} 
                \else
                    \definecolor{boxColor}{RGB}{224,224,160} 
                \fi
            \fi
            \ifnum\y=4
                \ifodd\x
                    \definecolor{boxColor}{RGB}{192,224,192} 
                \else
                    \definecolor{boxColor}{RGB}{224,192,192} 
                \fi
            \fi

            \fill[boxColor] (-0.43+\m, -\y-0.45) 
                            rectangle (0.43+\n, -\y+0.45);
                            
            \draw[dashed, thick, black] (-0.43+\m, -\y-0.45) 
                                       rectangle (0.43+\n, -\y+0.45);
        }
    }
    
    \foreach \y/\row in {0/green, 1/green, 2/mix, 3/mix, 4/mix} {
        \foreach \x in {0,...,15} {
            \ifnum\y=0
                \definecolor{nodeColor}{RGB}{0,255,0} 
            \fi
            \ifnum\y=1
                \definecolor{nodeColor}{RGB}{0,255,0} 
            \fi
            \ifnum\y=2
                \definecolor{nodeColor}{RGB}{0,255,0} 
                \ifnum\x<4
                    \definecolor{nodeColor}{RGB}{255,0,0} 
                \fi
                \ifnum\x>11
                    \definecolor{nodeColor}{RGB}{255,0,0} 
                \fi
            \fi
            \ifnum\y=3
                \definecolor{nodeColor}{RGB}{0,255,0} 
                \ifnum\x<6
                    \definecolor{nodeColor}{RGB}{255,0,0} 
                \fi
                \ifnum\x>9
                    \definecolor{nodeColor}{RGB}{255,0,0} 
                \fi
            \fi
            \ifnum\y=4
                \definecolor{nodeColor}{RGB}{0,255,0} 
                \ifnum\x<7
                    \definecolor{nodeColor}{RGB}{255,0,0} 
                \fi
                \ifnum\x>9
                    \definecolor{nodeColor}{RGB}{255,0,0} 
                \fi
            \fi

            \node[circle, draw=black, thick, fill=nodeColor, text=black, minimum size=7mm, inner sep=0pt, font=\bfseries] 
                (N\x-\y) at (\x, -\y) {\textbf{\x}};
        }
    }

    \draw[line width=1.2mm, blue, ->] (N0-0) -- (N0-1);
    \draw[line width=1.2mm, blue, ->] (N0-1) -- (N4-2);
    \draw[line width=1.2mm, blue, ->] (N4-2) -- (N6-3);
    \draw[line width=1.2mm, blue, ->] (N6-3) -- (N7-4);
    
    \draw[line width=1.2mm, purple, ->] (N15-0) -- (N15-1);
    \draw[line width=1.2mm, purple, ->] (N15-1) -- (N11-2);
    \draw[line width=1.2mm, purple, ->] (N11-2) -- (N9-3);
    \draw[line width=1.2mm, purple, ->] (N9-3) -- (N9-4);
\end{tikzpicture}

  \caption{The diagram illustrates the computationally efficient maximum likelihood method using binary search over possible intervals. The figure demonstrates the method for $N=16$, where $N$ represents the number of power measurements. The blue and purple lines respectively depict the trajectory of the detected start and end of the optimal interval across different stages. The signal, detected within the interval $X_{[7,9]}$, is highlighted by green circles.}
  \label{fig:ML_efficient_diagram}
\end{figure*}

At each iteration, it then iteratively
refines first the estimate for the left boundary
$\wh{a}^{(t)}$ and then the right boundary
$\wh{b}^{(t)}$.  The update for the left boundary
searches over candidates:
\begin{equation}
    \wh{a}^{(t+1)} \in \{ 
        \wh{a}^{(t)}-\ell_{t+1},\wh{a}^{(t)},
        \wh{a}^{(t)}+\ell_{t+1}\}
\end{equation}
so that it searches over one index to the right and left.  For each candidate, $\wh{a}^{(t+1)}$,
the algorithm evaluates the objective
$J([\wh{a}^{(t+1)}, \wh{b}^{(t)})$) 
where $J(S)$ is the likelihood in
\eqref{eq:Js_max}.  Similarly, the optimization
over the right boundary searches over values
for $\wh{b}^{(t+1)}$ in the set:
\begin{equation}
    \wh{b}^{(t+1)} \in \{ 
        \wh{b}^{(t)}-\ell_{t+1},\wh{b}^{(t)},
        \wh{b}^{(t)}+\ell_{t+1}\},
\end{equation}
so that it searches over one index to the left
and right.

\subsection{Complexity}

The algorithm~\ref{alg:dyad_search} comprises $\log_2(N)$ steps, within each of which the energy is assessed across $N/2^{m}$ intervals. Consequently, $O(N)$ computations are required in total to determine the energy of all intervals.
For Algorithm~\ref{alg:int_estimation},
there are two key computational savings.  
First, each evaluation
of $J(S)$ does not require recomputing the
average energy $\wb{X}_S$.  Recomputing $\wb{X}_S$
directly requires taking a sum over $|S|$
elements, which can be as large as $N$.
However, the algorithm uses the pre-computed
values $Z_{m,i}$ to compute $\wb{X}_S$.
Secondly, the search involves only 
$M=\log_2(N)$ iterations, and in each iteration,
there is a bounded number of operations.
Hence, the overall complexity of Algorithm~\ref{alg:int_estimation} is $O(\log_2(N))$. So, the proposed method, which is the union of the two algorithms, has an overall complexity of $O(N)$.


\section{Asymptotic Consistency of the Binary Search}
\label{sec:analysis}
In this section, we prove that the binary 
interval estimation algorithm, Algorithm~\ref{alg:int_estimation}, is asymptotically
consistent in the limit of
large $N$ under a fixed \ac{SNR} per bin $\gamma$.
Specifically, 
we consider the $d=1$ dimensional case and
examine a sequence of
problems indexed by the total length $N$.
We fix a ``true" \ac{SNR} $\gamma$ and assume that
the true signal
interval
\begin{equation} \label{eq:Struelim}
    S^0_N= [a^0_N, b^0_N), \quad a^0_N = \lfloor{\alpha_0 N} \rfloor, \quad 
    b^0_N = \lfloor{\beta_0 N} \rfloor,
\end{equation}
for some constants $0 < \alpha_0 < \beta_0 < 1$
that do not depend on $N$.
Assume that $X_n$, $n=0,\ldots,N$
are generated i.i.d. from the model
\eqref{eq:xscalar}.
We let $\wh{a}^{(t)}_N, \wh{b}^{(t)}_N$
be the outputs of Algorithm~\ref{alg:int_estimation},
and let $\wh{S}^{(t)}_N$ denote the corresponding
interval:
\begin{equation} \label{eq:Sest}
    \wh{S}^{(t)}_N= [\wh{a}^{(t)}_N, \wh{b}^{(t)}_N)
\end{equation}
A common metric for the performance
of the algorithm is the so-called 
intersection over union, or IoU:
\begin{equation} \label{eq:iout}
\mathrm{IoU}^{(t)}_N = \frac{|\wh{S}^{(t)}_N \cap S^0_N|}{|\wh{S}^{(t)}_N \cup S^0_N|}.
\end{equation}
Although the true interval is modeled
as deterministic, the estimated interval
$\wh{S}^{(t)}_N$ and, hence, the performance
metric $\mathrm{IoU}^{(t)}_N$ are random.

\begin{theorem} \label{thm:iou}
Under the above assumptions
\begin{equation} \label{eq:ioubnd}
    \lim_{N, t \rightarrow \infty} \mathrm{IoU}^{(t)}_N = 1
\end{equation}
almost surely.
\end{theorem}
\begin{proof} See Appendix~\ref{sec:proof_iou}
\end{proof}

This asymptotic consistency result (Theorem~\ref{thm:iou}) is important because it establishes that the proposed binary interval estimation algorithm reliably identifies the correct signal interval in the limit of large measurement sizes. Practically, this means that as more data is collected (larger $N$) and the algorithm is allowed sufficient refinement steps (larger $t$), the estimated interval converges exactly to the signal region identified by the exhaustive algorithm, guaranteeing high accuracy in signal detection tasks.

\section{Baseline Neural Networks} \label{sec:neural_networks}

It is important to compare the proposed algorithm to widely-used U-Net networks.  The U-Net architecture is widely used in segmentation and detection tasks~\cite{ronneberger2015u,siddique2021u,oktay2018attention}, and has also been proposed for spectral segmentation and signal detection~\cite{nguyen2024enhancing,uvaydov2024stitching,dakic2024spiking,park2022target,west2021wideband}.  Our simulations below will show that the proposed method offers both
improved complexity and performance compared to U-Nets.  

We selected the \ac{CNN}-based U-Net architecture as our primary baseline because it remains the de facto standard for semantic segmentation in signal processing, effectively capturing the local coherence required to detect contiguous signal intervals. While newer architectures, such as Transformers or hybrid CNNs-Transformers, excel in tasks requiring long-range dependency modeling, they typically incur significantly higher computational costs—often scaling quadratically ($O(N^2)$) due to self-attention mechanisms—compared to the efficiency of \acp{CNN}. Given that the proposed binary search method already operates near the theoretical detection bound with linear $O(N)$ complexity, employing a heavier Transformer baseline would primarily widen the computational gap without yielding statistically significant performance gains over the theoretical limit. Thus, the U-Net serves as a robust and sufficient benchmark to illustrate the trade-off between deep learning complexity and our efficient algorithmic approach.

\begin{figure*}[]
  \centering
  \includegraphics[width=0.8\textwidth]{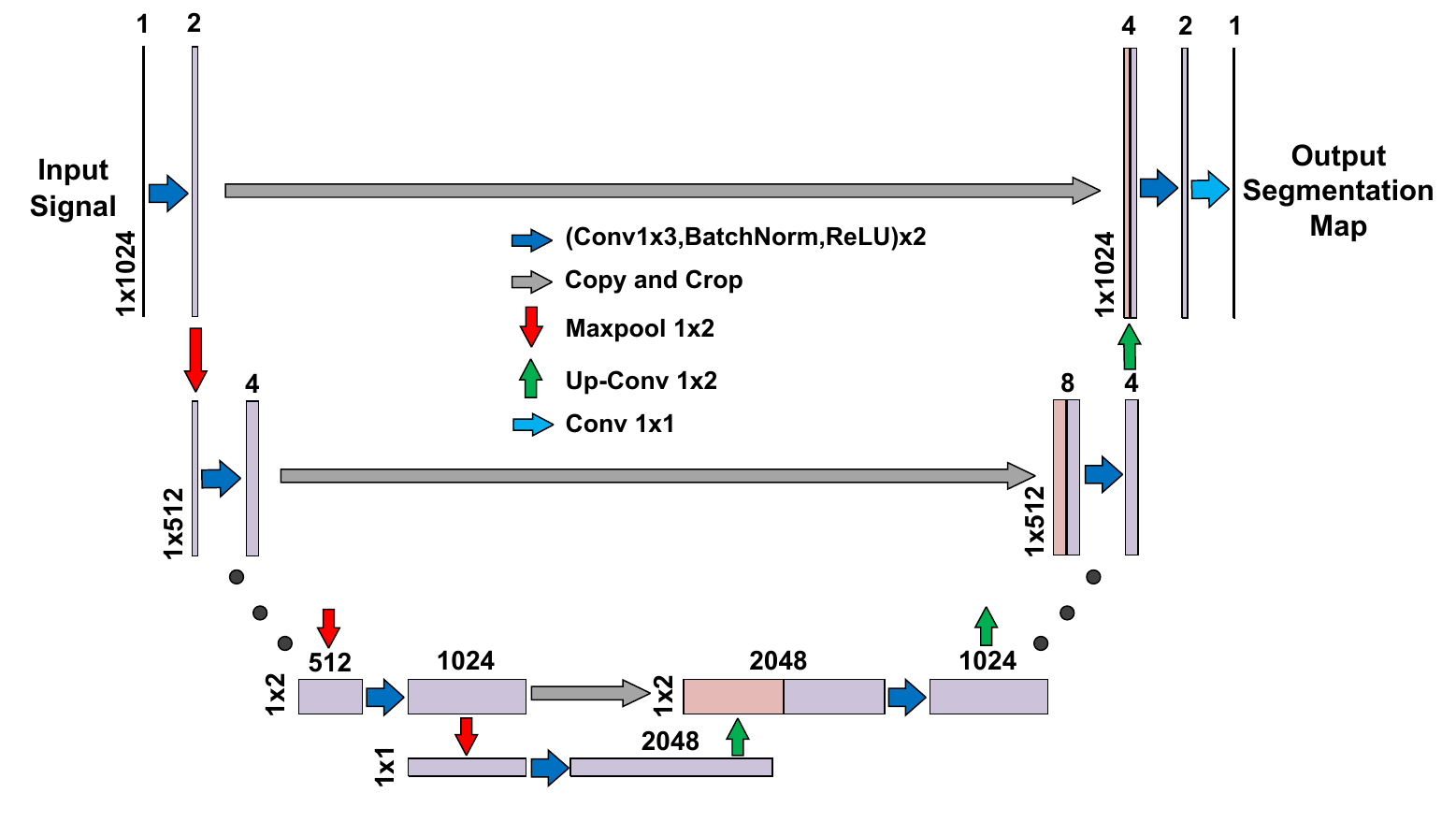}
    \caption{The architecture of the baseline U-Net used for signal detection in this paper. This architecture is designed for one-dimensional data. However, the same principles and methods can be extended to handle higher-dimensional data.
    For each feature map, dimensions are annotated at the bottom left, while the number of channels is indicated at the top. As illustrated in the figure, the process initiates with the input signal, upon which a convolutional block comprising two stages of $1\times3$ convolution, Batch Normalization, and ReLU activation is applied. This is followed by a Max Pooling operation, which is iteratively repeated until achieving a $1\times 1$ feature map with 2048 channels. The decoding phase commences with up-convolutions, wherein feature maps are concatenated with the corresponding encoder maps, applying the same convolutional block utilized during the encoding stage. Finally, a Conv$1\times1$ layer is employed to generate the output segmentation map.}
  \label{fig:unet}
\end{figure*}

For the one-dimensional case, we consider the U-Net depicted in Fig.~\ref{fig:unet}. For higher-dimensional data, the standard U-Net architecture is employed, with a sufficient number of encoding and decoding stages and appropriately scaled data and filter sizes.
In each stage, a two-step convolutional block is applied. This block consists of a convolution operation, followed by Batch Normalization and ReLU activation, and is concluded with a Max-Pooling layer to downsample the signal by a factor of two. During the encoding stages, feature map information is progressively captured across multiple channels, culminating in $M$ channels of $1 \times 1$ data.
This process parallels the proposed binary search method. Initially, $J(S)$ is computed over the entire interval, similar to the first stage of the U-Net encoder. The search is then refined over progressively smaller intervals, analogous to the subsequent encoder stages of the U-Net, until the signal is examined in each bin corresponding to the final stage of the U-Net encoder.

In the decoding phase, similar to the conventional U-Net architecture, we iteratively up-sample the feature maps and concatenate them with the corresponding encoder feature maps. A similar convolutional block is then applied to the concatenated feature maps. This process continues until we obtain a two-channel output segmentation map, where a Conv$1 \times 1$ layer is used to generate the final segmentation mask. Intuitively, the decoder seeks to reconstruct the signal interval information from the encoded channels using convolutional blocks, which parallels the functionality of our binary search algorithm.


\section{Experimental Results} \label{sec:experiments}

Our goal is to compare the detection accuracy and signal set estimation error of the proposed binary search method to more complex exhaustive \ac{ML}. Additionally, for the signals that are detected, we will compare the estimation error to a trained U-Net.  
We consider two scenarios for these comparisons:
\begin{itemize}
    \item \emph{1D detection}:  We take $N=1024$ points.  When a signal is present, it is in an unknown interval with variable length $\ell$ and variable \ac{SNR}.
    \item \emph{2D detection}:  We take $N=128 \times 128$ points.  When a signal is present, it is on an unknown square of size $\ell \times \ell$ and variable \ac{SNR}.
\end{itemize}

Specifically, the received power measurements $X_n$ are modeled as independent exponential random variables with a mean $\Exp[X_n] = 1 + \gamma$ if the bin $n$ belongs to the signal region $S$, and $\Exp[X_n] = 1$ otherwise. This corresponds to a standard non-coherent energy detection setup, where $\gamma$ represents the \ac{SNR}.
We train U-Nets for each of the two cases.  In both cases, we generate a dataset of $n=200,000$ samples.
Since the U-Net is only used for the estimation of the signal set, all $n$ samples have a signal present.  Hence, each sample $i$ can be represented as a pair $(X_i,S_i)$, where $X_i$ is the vector
or array of power values, and $S_i$ is the true signal interval.  We vary the \ac{SNR} in the samples from \SI{-3}{dB} to \SI{20}{dB}, with an emphasis on lower \acp{SNR}.  For the 1D case, the interval size is varied
from 1 to 256, and in the 2D case $\ell$ is varied from 1 to 64. The dataset is divided into training and test sets with an 80/20\% ratio. A summary of the hyperparameters and settings used for the training and evaluation of the U-Net is provided in Table.~\ref{tab:unet_params}.

\begin{table}[t]
\caption {Summary of hyperparameters and settings used for U-Net training and evaluation.}
\centering
\vspace{0.1cm}
\footnotesize
\renewcommand{\arraystretch}{1.1}
\begin{tabular}{>{\arraybackslash}m{3.2cm}||>{\centering\arraybackslash}m{3.5cm}}

    \hline
    {\normalsize \textbf{Hyperparameter}} & {\normalsize \textbf{Value}}
    \tabularnewline \hline \hline
    Optimization Algorithm & Adam
    \tabularnewline \hline
    Initial Learning Rate & $10^{-2}$
    \tabularnewline \hline
    Learning Rate Scheduler & StepLR
    \tabularnewline \hline
    Batch Size & 64
    \tabularnewline \hline
    Number of Training Epochs & 50
    \tabularnewline \hline
    Training Loss Function & Binary Cross-Entropy (BCE)
    \tabularnewline \hline
    Evaluation Metric & Intersection over Union (IoU)
    \tabularnewline \hline
    U-Net Parameters (1D) & 43,380,007
    \tabularnewline \hline
    U-Net Parameters (2D) & 1,949,011
    \tabularnewline \hline
    PyTorch Version & 2.6.0
    \tabularnewline
    \hline
    
\end{tabular}
\label{tab:unet_params}
\end{table}

\begin{figure*}[htbp]
    \centering
    
    
    \begin{subfigure}[t]{1.0\textwidth}
        \centering
        \includegraphics[width=\textwidth]{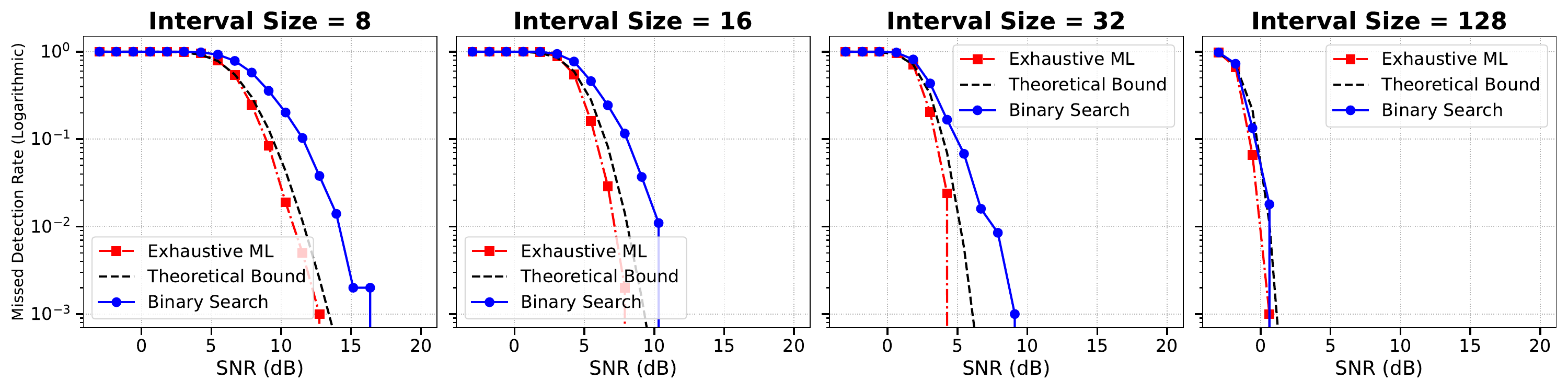}
        \caption{The missed detection rate of maximum likelihood methods for different \acp{SNR} (false alarm rate fixed at $10^{-6}$) for the one-dimensional data.}
        \label{fig:ss_sw_snr_missed_rate_1d}        
    \end{subfigure}

    \begin{subfigure}[t]{1.0\textwidth}
        \centering
        \includegraphics[width=\textwidth]{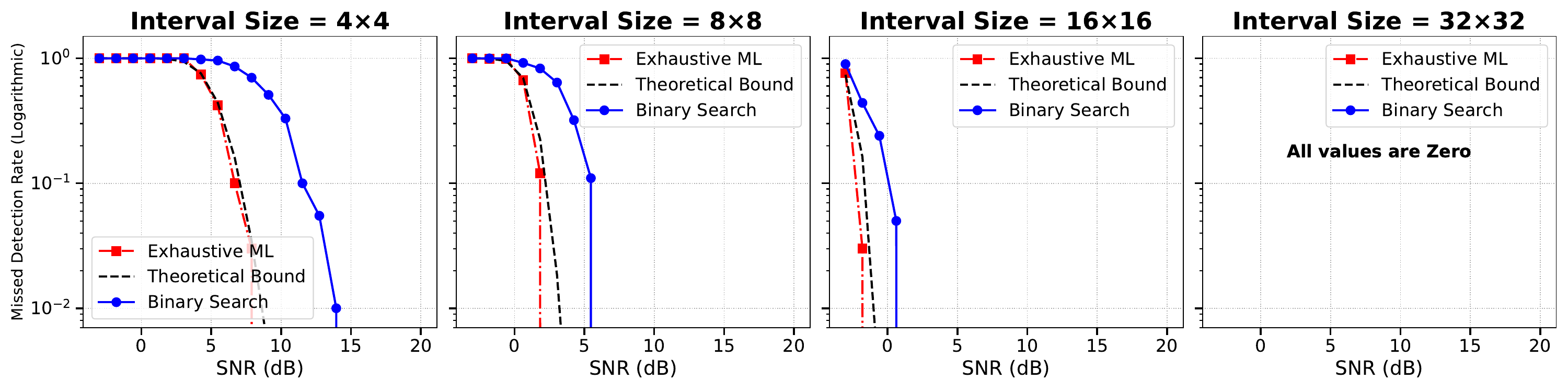}
        \caption{The missed detection rate of maximum likelihood methods for different \acp{SNR} (false alarm rate fixed at $10^{-6}$) for the two-dimensional data.}
        \label{fig:ss_sw_snr_missed_rate_2d}        
    \end{subfigure}
    
    \caption{The missed detection rate of the exhaustive and binary search maximum likelihood estimations and U-Net for 1D and 2D signals with \acp{SNR} ranging from -3 to 20 dB. Each plot represents the performance for a specific fixed signal size. As expected, increasing the \ac{SNR} improves the performance of both methods.}
    \label{fig:ss_sw_snr_missed}
    
\end{figure*}

\begin{figure*}[htbp]
    \centering
    
    \begin{subfigure}[t]{1.0\textwidth}
        \centering
        \includegraphics[width=\textwidth]{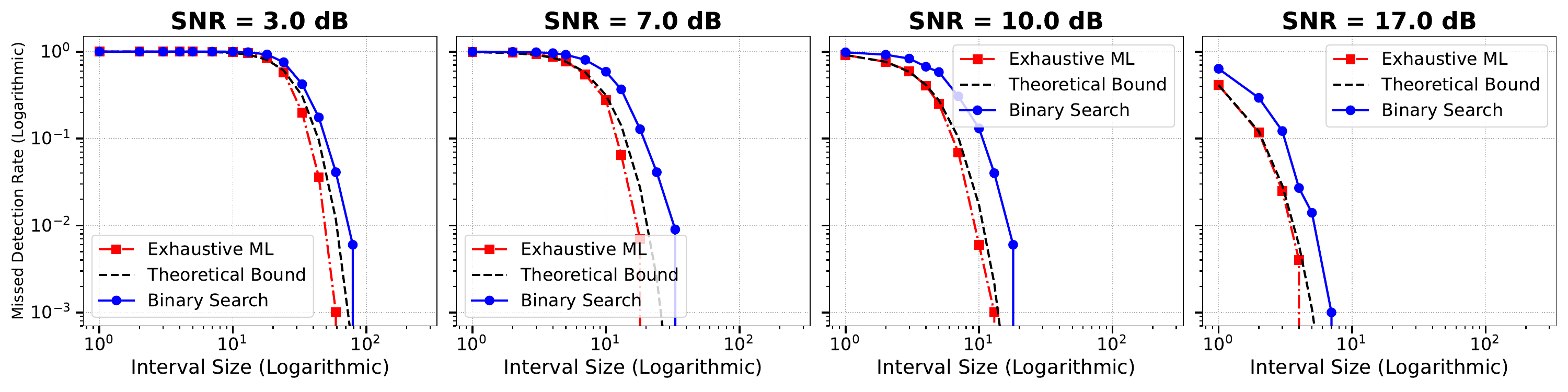}
        \caption{The missed detection rate of maximum likelihood methods for different signal sizes (false alarm rate fixed at $10^{-6}$) for the one-dimensional data.}
        \label{fig:ss_sw_size_missed_rate_1d}
    \end{subfigure}

    \begin{subfigure}[t]{1.0\textwidth}
        \centering
        \includegraphics[width=\textwidth]{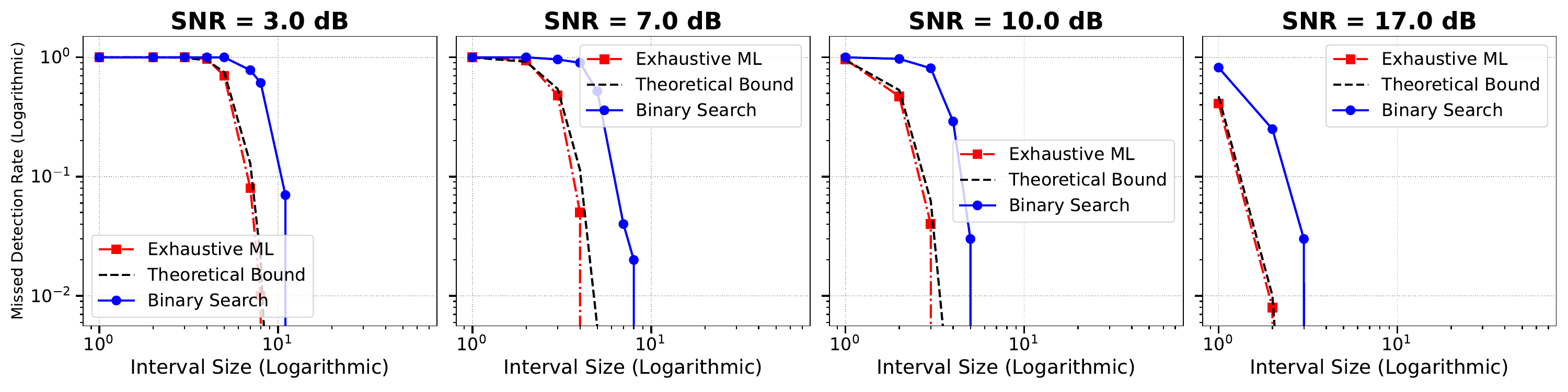}
        \caption{The missed detection rate of maximum likelihood methods for different signal sizes (false alarm rate fixed at $10^{-6}$) for the two-dimensional data.}
        \label{fig:ss_sw_size_missed_rate_2d}
    \end{subfigure}
    
    \caption{The missed detection rate of the exhaustive and binary search maximum likelihood estimations and U-Net for 1D and 2D signals with sizes ranging from 1 to 256 (out of \(N=1024\)) for 1D and 1 to 64 (out of \(N=128\)) for 2D. Each plot represents the performance for a specific fixed signal \ac{SNR}.}
    \label{fig:ss_sw_size_missed}
    
\end{figure*}

For detection accuracy, we fix the threshold for a false alarm rate of $10^{-6}$ using the bound in Lemma~\ref{lem:pfa}. For every candidate interval, the corresponding threshold $t_{\ell}$ is calculated and subsequently employed within the \ac{GLRT} detector.
In our experiments, we adopt a stringent false alarm probability of $P_{FA} = 10^{-6}$ to reflect the requirements of practical spectrum sensing and adversarial detection scenarios, where false alarms can incur significant operational costs. This conservative setting intentionally imposes a higher detection threshold, thereby creating a challenging test environment for the proposed method. As will be detailed later, despite this stringent constraint, the proposed binary-search-based detector exhibits consistently high detection performance, demonstrating its robustness and effectiveness under low-SNR and high-uncertainty conditions.
Figures.~\ref{fig:ss_sw_snr_missed} and \ref{fig:ss_sw_size_missed} show the missed detection rates for
exhaustive \ac{ML} and binary search methods. We observe that in the 1D case, the detection is no worse than \SI{3}{dB} and 
no worse than \SI{6}{dB} in the 2D case. This difference is explicable since, in the worst
case, the binary search captures $1/2$ the energy in the 1D case and $1/4$ the energy in the 2D case.
Importantly, as the interval sizes increase, the gap decreases, indicating that the proposed method
has minimal loss for larger interval sizes.

\begin{figure*}[htbp]
    \centering
    
    \begin{subfigure}[t]{1.0\textwidth}
        \vspace{0pt}
        \centering
        \includegraphics[width=\textwidth]{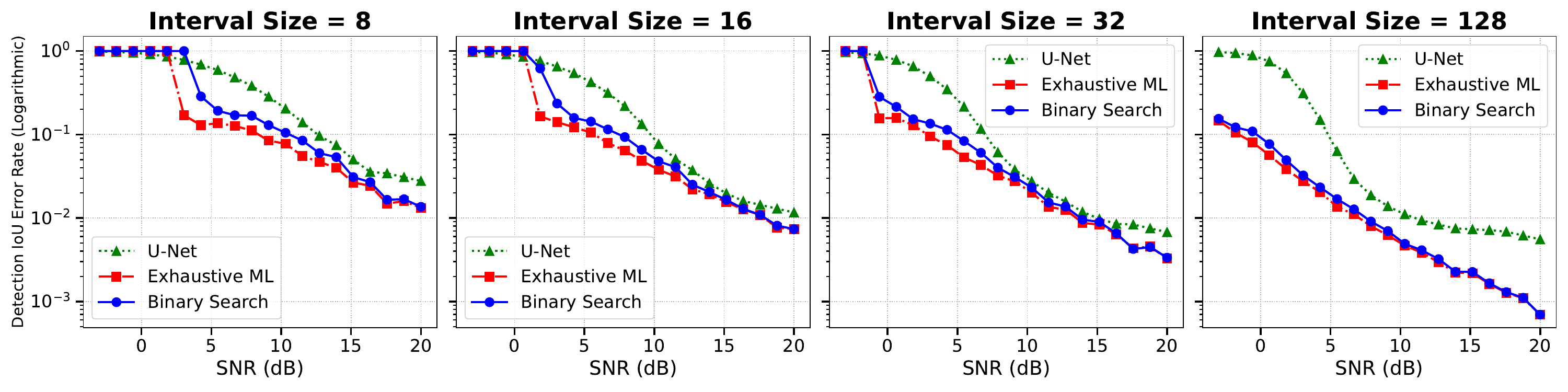}
        \caption{The detection \ac{IOU} error rate of different methods for different \acp{SNR} (false alarm rate fixed at $10^{-6}$) for the one-dimensional data.}
        \label{fig:ss_sw_snr_det_rate_1d}
    \end{subfigure}

    \begin{subfigure}[t]{1.0\textwidth}
        \vspace{0pt}
        \centering
        \includegraphics[width=\textwidth]{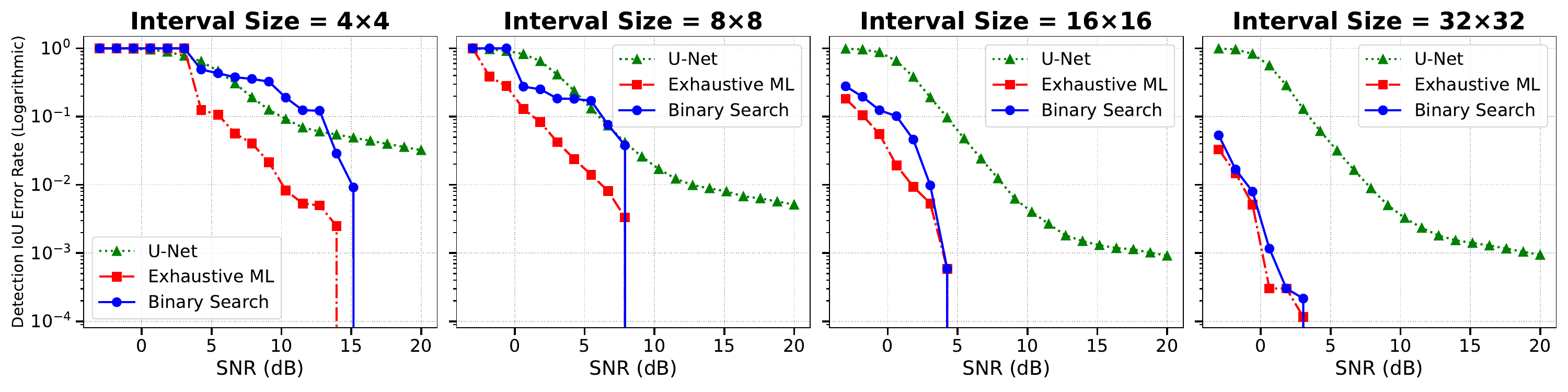}
        \caption{The detection \ac{IOU} error rate of different methods for different \acp{SNR} (false alarm rate fixed at $10^{-6}$) for the two-dimensional data.}
        \label{fig:ss_sw_snr_det_rate_2d}
    \end{subfigure}
    
    \caption{The detection \ac{IOU} error rate of the exhaustive and binary search maximum likelihood estimations and U-Net for 1D and 2D signals with \acp{SNR} ranging from -3 to 20 dB. Each plot represents the performance for a specific fixed signal size. As expected, increasing the SNR improves the performance of both methods. The figures indicate that the performance of the two methods is nearly identical for larger signal sizes. For smaller signal sizes, the binary search method still performs similarly to the exhaustive \ac{ML} at low and high \acp{SNR}, with only a slight, non-significant performance drop observed at intermediate \acp{SNR}. In most cases, the binary search \ac{ML} method surpasses the U-Net in the \ac{IOU} metric while maintaining significantly lower computational complexity. The performance is generally better for 2D data compared to 1D data.}
    \label{fig:ss_sw_snr_det}
    
\end{figure*}

\begin{figure*}[htbp]
    \centering
    
    \begin{subfigure}[t]{1.0\textwidth}
        \vspace{0pt}
        \centering
        \includegraphics[width=\textwidth]{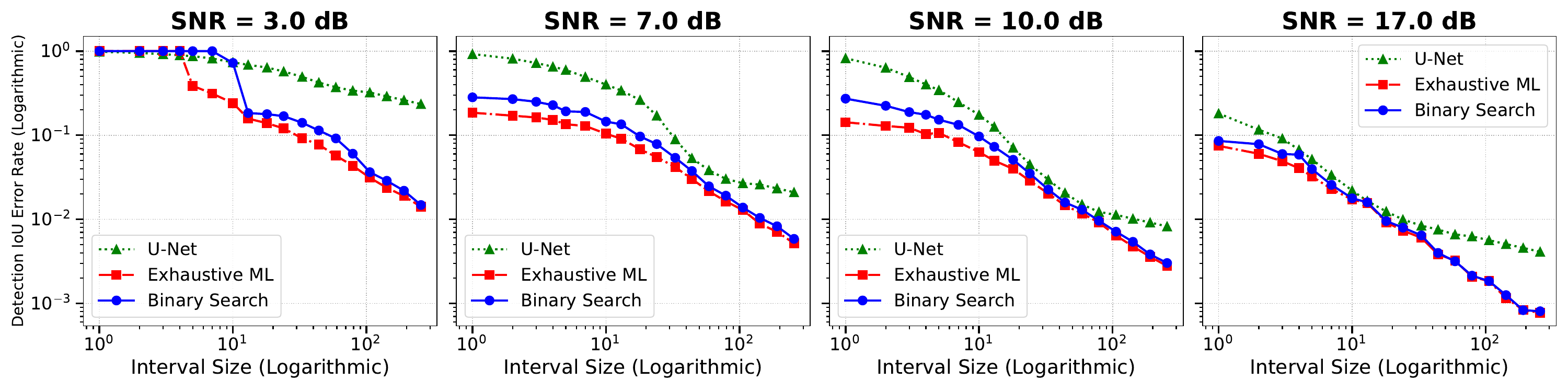}
        \caption{The detection \ac{IOU} error rate of different methods for different signal sizes (false alarm rate fixed at $10^{-6}$) for the one-dimensional data.}
        \label{fig:ss_sw_size_det_rate_1d}
    \end{subfigure}

    \begin{subfigure}[t]{1.0\textwidth}
        \vspace{0pt}
        \centering
        \includegraphics[width=\textwidth]{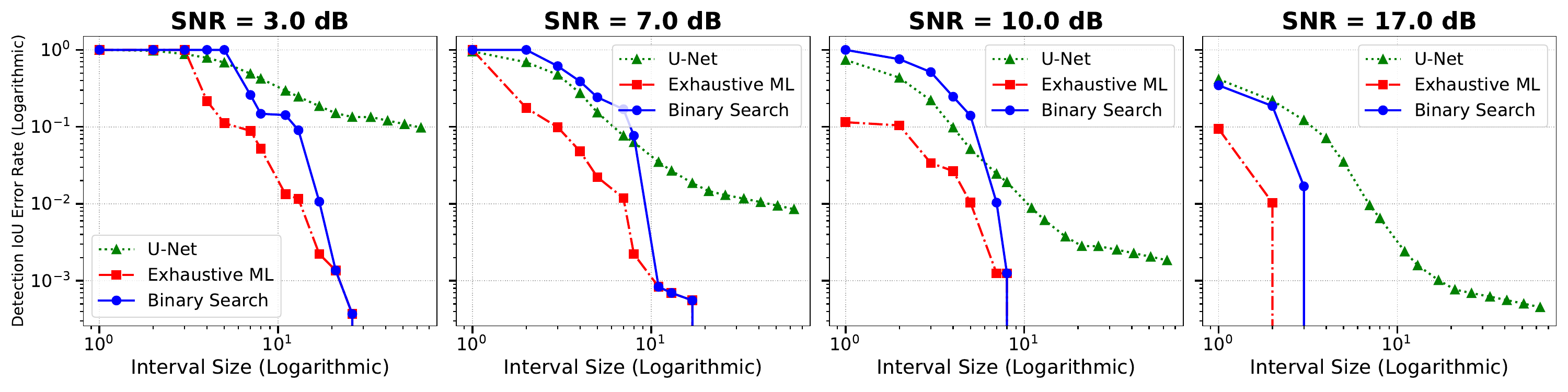}
        \caption{The detection \ac{IOU} error rate of different methods for different signal sizes (false alarm rate fixed at $10^{-6}$) for the two-dimensional data.}
        \label{fig:ss_sw_size_det_rate_2d}
    \end{subfigure}
    
    \caption{The detection \ac{IOU} error rate of the exhaustive and binary search maximum likelihood estimations and U-Net for 1D and 2D signals with sizes ranging from 1 to 256 (out of \(N=1024\)) for 1D and 1 to 64 (out of \(N=128\)) for 2D. Each plot represents the performance for a specific fixed signal \ac{SNR}. Increasing the signal size improves the performance of both methods, as detecting smaller signals is more challenging. As shown in the figure, the binary search method exhibits a slight, non-significant performance drop, which is less pronounced for larger \acp{SNR}. Similar to Fig.~\ref{fig:ss_sw_snr_det}, the performance of both methods is much closer for very small and very large signal sizes. Similarly in most cases, the binary search \ac{ML} method outperforms the U-Net in the \ac{IOU} metric while keeping lower computational complexity. The performance is generally better for 2D data compared to 1D data.}
    \label{fig:ss_sw_size_det}
    
\end{figure*}

Figures.~\ref{fig:ss_sw_snr_det} and \ref{fig:ss_sw_size_det} show the estimation accuracy for the detected signals. 
 Here, we compare the exhaustive \ac{ML}, binary search, and U-Net.  For all three, we plot the detection \ac{IOU} error rate, defined as:
\begin{equation}
    \text{IOU Error Rate}= 1 - \frac{S \cap \hat{S}}{S \cup \hat{S}}
\end{equation}
 where $S$ represents the ground-truth interval and $\hat{S}$ corresponds to the estimated interval. From the figures, we see that for any \ac{SNR}, as the interval sizes increase, the detection error rate approaches zero, as predicted by Theorem~\ref{thm:iou}.  Second, for larger interval sizes, the gap between the exhaustive \ac{ML} and binary search decreases, again suggesting that the proposed method will perform well for larger intervals.
 
\begin{figure}
    \centering
    \includegraphics[width=1\linewidth]{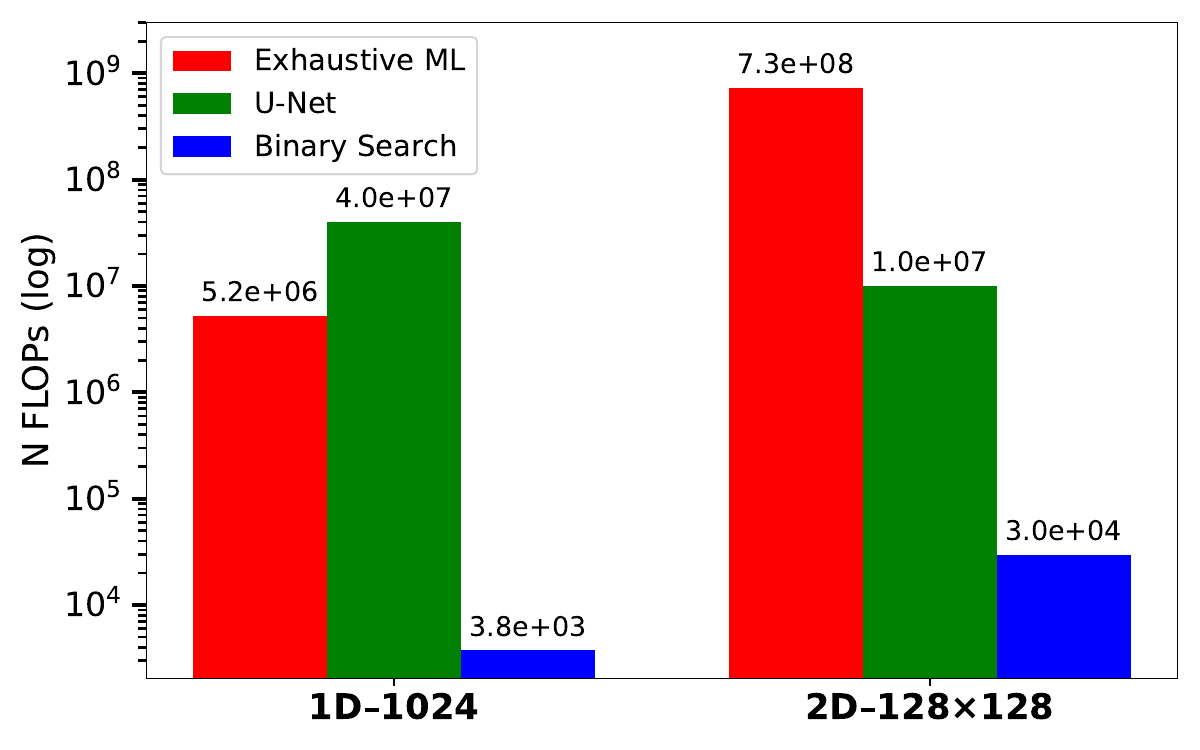}
    \caption{Comparison of number of \acp{FLOP} needed for the proposed binary search method, exhaustive \ac{ML} method, and U-Net for one-dimensional signals of size $1024$ and two-dimensional signals of size $128 \times 128$. The results indicate that the computational complexity of the proposed method is reduced by three to four orders of magnitude.}
    \label{fig:flops}
\end{figure}

Finally, the proposed method significantly outperforms U-Net in most interval sizes. This improvement in performance occurs even though the proposed binary search method is significantly simpler.  
As a way to compare, we can count the number of operations for detection (Algorithm.~\ref{alg:dyad_search}), binary interval estimation (Algorithm.~\ref{alg:int_estimation}), exhaustive \ac{ML}, and U-Net. Table.~\ref{tab:n_flops} and Fig.~\ref{fig:flops} present a comprehensive summary of the floating point operations required for each algorithm considered. These figures were estimated using a Python-based simulator (utilizing standard NumPy, SciPy, and PyTorch libraries). As evidenced by the data in the table, the proposed binary search algorithm requires significantly fewer operations—by several orders of magnitude—compared to both U-Net and exhaustive \ac{ML} methods, applicable to both 1D and 2D scenarios.

\begin{table}[t]
\caption {Comparison of number of \acp{FLOP} needed for the proposed binary search method, exhaustive \ac{ML} method, and U-Net}
\centering
\vspace{0.1cm}
\footnotesize
\renewcommand{\arraystretch}{1.1}
\begin{tabular}{>{\arraybackslash}m{2.8cm}||>{\centering\arraybackslash}m{1.8cm}|>{\centering\arraybackslash}m{1.8cm}}

    \hline
    \multirow{2}{*}{\normalsize \textbf{Algorithm}} & \multicolumn{2}{c}{\normalsize \textbf{Number of \acp{FLOP}}} \\
    \cline{2-3}
     & \textbf{$1D\!-\!1024$} & \textbf{$2D\!-\!128 \!\times\! 128$} \\
    \hline \hline

    Exhaustive \ac{ML} & $5.24 \times 10^{6}$ & $7.27 \times 10^{8}$
    \tabularnewline \hline
    U-Net & $4.0 \times 10^{7}$ & $10^{7}$
    \tabularnewline \hline
    \multicolumn{3}{l}{\textbf{Binary Search (Proposed Method)}}
    \tabularnewline \hline
    Algorithm~\ref{alg:dyad_search} & $3.13 \times 10^{3}$ & $2.86 \times 10^{4}$
    \tabularnewline \hline
    Algorithm~\ref{alg:int_estimation} & $6.2 \times 10^{2}$ & $9.1 \times 10^{2}$
    \tabularnewline \hline
    \textbf{Binary Search-Total} & $\mathbf{3.75 \times 10^{3}}$ & $\mathbf{2.95 \times 10^{4}}$
    \tabularnewline \hline

\end{tabular}
\label{tab:n_flops}
\end{table}

\subsection{Effect of Noise Estimation} \label{sec:noise_est_effect}
\begin{figure*}[htbp]
    \centering

    \begin{subfigure}[t]{1.0\textwidth}
        \vspace{0pt}
        \centering
        \includegraphics[width=\textwidth]{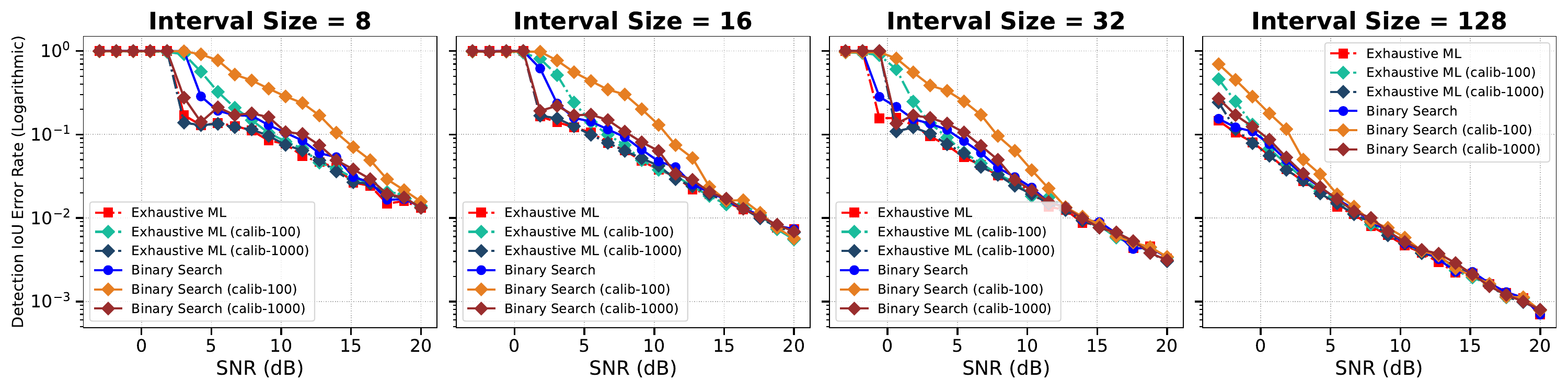}
        \caption{The detection \ac{IOU} error rate of different methods for a range of SNRs (false alarm rate fixed at $10^{-6}$) for the one-dimensional data.}
        \label{fig:ss_sw_snr_det_rate_1d_calib}
    \end{subfigure}
    
    \begin{subfigure}[t]{1.0\textwidth}
        \vspace{0pt}
        \centering
        \includegraphics[width=\textwidth]{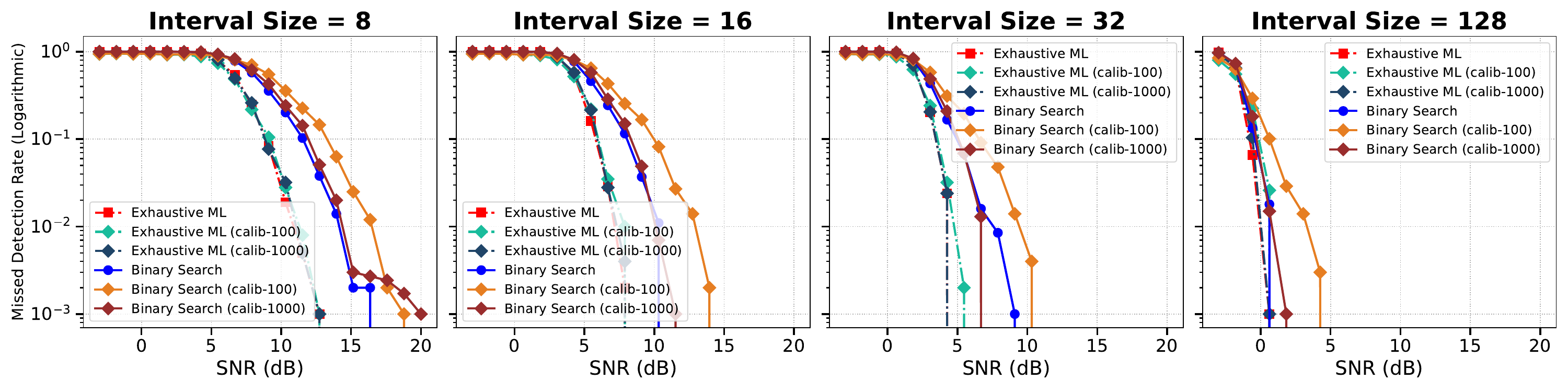}
        \caption{The missed detection rate of maximum likelihood methods for a range of SNRs (false alarm rate fixed at $10^{-6}$) for the one-dimensional data.}
        \label{fig:ss_sw_snr_missed_rate_1d_calib}
    \end{subfigure}

    \caption{The detection \ac{IOU} error rate and missed detection rate of the exhaustive and binary search maximum likelihood estimations for 1D signals across \acp{SNR} ranging from -3 to 20 dB. Each curve corresponds to a fixed signal size. Results are shown for both the ideal case with perfectly known noise variance and the non-ideal case where the noise power is estimated from $N_{\mathrm{ref}}$ calibration samples ($N_{\mathrm{ref}}$ is indicated within parentheses). Increasing $N_{\mathrm{ref}}$ reduces the estimation variance and improves detection performance, illustrating the trade-off between calibration overhead and detection reliability.}
    \label{fig:ss_sw_snr_calib}

\end{figure*}

Up to this point, we have implicitly assumed that the noise variance is perfectly known and normalized to unity. In practical systems, however, the noise power must be estimated during a calibration phase, and any estimation error can impact subsequent detection performance.
To quantify the effect of imperfect noise estimation, we conduct the following experiment. Let $N_{\mathrm{ref}}$ denote the number of degrees of freedom allocated for noise power estimation. During the calibration, it is assumed that no signals are present and that the received samples consist purely of noise. The noise variance is then estimated as
\(\widehat{\sigma}^2 = \frac{1}{N_{\mathrm{ref}}} \sum_{i=1}^{N_{\mathrm{ref}}} |w_i|^2,\)
where ($w_i \sim \mathcal{CN}(0, \sigma^2)$ ) are independent noise samples. This estimator follows a scaled chi-squared distribution with ($2N_{\mathrm{ref}}$) degrees of freedom, implying that its relative variance decreases as ($1/N_{\mathrm{ref}}$).
Subsequently, in the detection phase, the received samples are normalized using the estimated noise power, i.e.,
\(X_n^{'} = \frac{X_n}{\widehat{\sigma}},\)
Where $X_n$ represents the received samples, and $X_n^{'}$ denotes the normalized samples, respectively. The detection statistic is then computed based on $X_n^{'}$, assuming unity noise variance, even though the true noise level may differ due to estimation error.
We evaluate the resulting detection performance by plotting the detection \ac{IOU} error rate and the probability of missed detection ($P_{\subsf MD}$) versus \ac{SNR} for different values of $N_{\mathrm{ref}}$, as illustrated in Fig.~\ref{fig:ss_sw_snr_calib}. This analysis highlights the trade-off between calibration overhead and detection accuracy: a larger ($N_{\mathrm{ref}}$) yields more reliable noise estimates and better detection performance, whereas a smaller ($N_{\mathrm{ref}}$) leads to higher variance in the estimated noise power and degraded performance.

\subsection{Performance Comparison With an Oracle Interval Detector} \label{sec:known_interval_effect}
\begin{figure*}[htbp]
    \centering
    
    \begin{subfigure}[t]{1.0\textwidth}
        \centering
        \includegraphics[width=\textwidth]{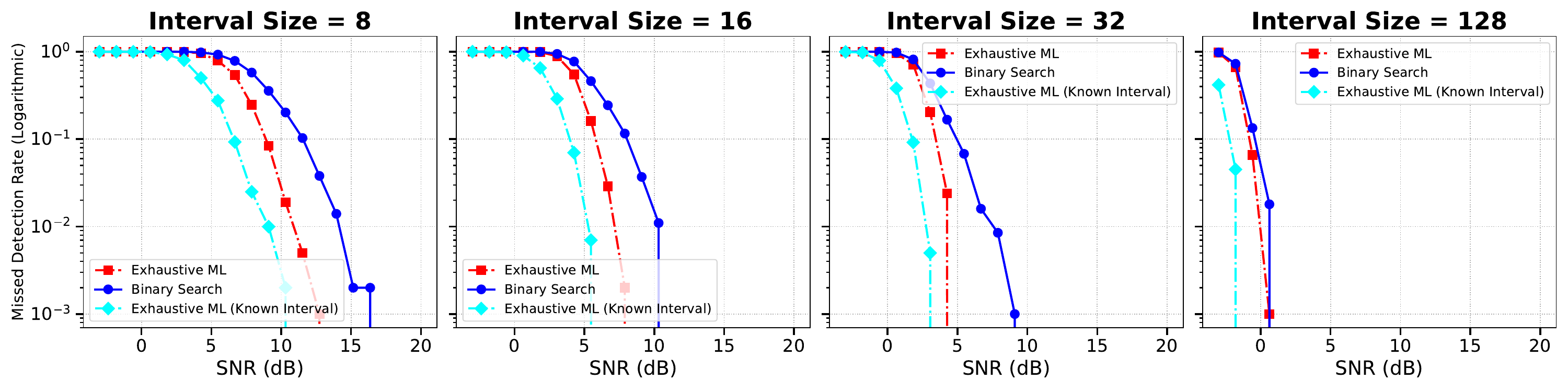}
        \caption{The missed detection rate of maximum likelihood methods for different \acp{SNR} (false alarm rate fixed at $10^{-6}$) for the one-dimensional data.}
        \label{fig:ss_sw_snr_missed_rate_1d_known}        
    \end{subfigure}

    \begin{subfigure}[t]{1.0\textwidth}
        \centering
        \includegraphics[width=\textwidth]{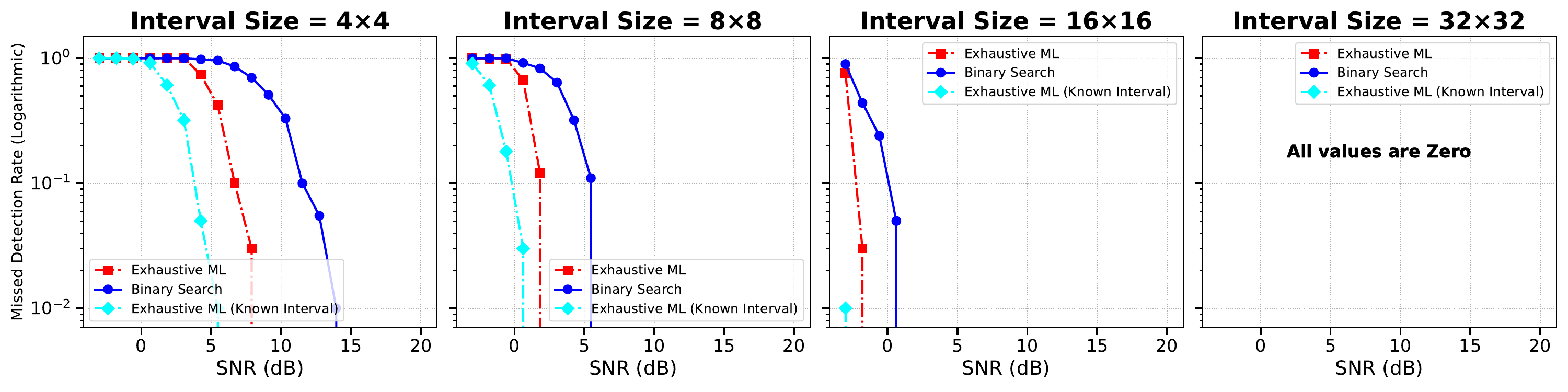}
        \caption{The missed detection rate of maximum likelihood methods for different \acp{SNR} (false alarm rate fixed at $10^{-6}$) for the two-dimensional data.}
        \label{fig:ss_sw_snr_missed_rate_2d_known}        
    \end{subfigure}
    
    \caption{The missed detection rate of the exhaustive and binary search detectors, for 1D and 2D signal cases across \acp{SNR} ranging from –3 dB to 20 dB. Each subplot corresponds to a fixed signal interval size $\ell_0$. The oracle (known interval) benchmark illustrates the lower bound on the missed detection rate achievable when the true signal interval is known. The results highlight the degradation introduced by interval uncertainty.}
    \label{fig:ss_sw_snr_missed_known}
    
\end{figure*}

The key challenge addressed in this work is that the signal is unknown and must be estimated jointly with detection. To illustrate the resulting performance impact, we compare the detector performance when the signal interval is not known with that of an oracle detector that has perfect knowledge of the true interval. Accordingly, we repeat the experiments described in Section~\ref{sec:experiments}, now including the oracle detector as a benchmark to assess the role of interval knowledge ($S$), as discussed in Section~\ref{sec:known_interval}. The results, shown in Fig.~\ref{fig:ss_sw_snr_missed_known}, indicate that when the signal interval is known a priori (oracle case), the detector achieves substantially better performance, yielding a lower missed-detection probability ($P_{\subsf{MD}}$) at a fixed false-alarm rate ($P_{\subsf{FA}}^0$). In contrast, the proposed \ac{GLRT} detector—which must infer the signal interval—incurs a moderate performance loss due to this additional uncertainty. This comparison quantitatively highlights the degradation introduced by interval uncertainty and confirms the theoretical trade-off between detection sensitivity and false-alarm control.

\subsection{Performance Analysis}
To rigorously evaluate the proposed method, we selected comparison baselines that bound performance from both theoretical and data-driven perspectives. The Exhaustive Maximum Likelihood method serves as the theoretical upper bound for detection performance. As derived in Lemma~\ref{lem:likelihood}, the Exhaustive \ac{ML} constitutes the optimal detector for the unknown interval problem under the exponential model. By benchmarking against this optimal standard, we quantify the precise performance trade-off incurred by the computational efficiency of our approach. The results in Figures.~\ref{fig:ss_sw_snr_missed} and \ref{fig:ss_sw_size_missed} demonstrate that the proposed Binary Search algorithm maintains performance within 3 dB of this theoretical limit (in the 1D case), validating its near-optimal efficacy.

Conversely, the U-Net architecture was selected to represent state-of-the-art, high-complexity \ac{DL} approaches. While \ac{DL} methods are increasingly popular for spectral segmentation, comparing them against U-Net highlights that our statistically derived \ac{GLRT} approach achieves superior \ac{IOU} and detection rates (Figures.~\ref{fig:ss_sw_snr_det} and \ref{fig:ss_sw_size_det}) without the massive computational overhead (Table~\ref{tab:n_flops}) or the extensive training data requirements inherent to neural networks. We prioritized these baselines over wavelet-based or edge-detection techniques, as the latter often rely on signal smoothness assumptions that may not hold for the non-coherent power measurements considered here, and typically incur higher computational complexity ($O(M.N \log N)$ where $M$ is the number of analyzed scales) compared to the $O(N)$ efficiency of our proposed method.

The superior performance of the proposed estimator in terms of \ac{FAP}, and detection probability stems from two fundamental properties of the \ac{GLRT}:
\begin{itemize}
    \item Analytically Controlled \ac{FAP}: Unlike heuristic edge detectors or neural networks, where the \ac{FAP} is often empirical, our method ensures a strict Constant False Alarm Rate. As established in Lemma~\ref{lem:pfa}, we derive a closed-form bound for the false alarm probability based on the Chi-squared distribution. This allows for the precise calculation of thresholds $u_l$ required to meet a specific target $P_{FA}$ (e.g., $10^{-6}$). This rigorous statistical control minimizes false positives more effectively than learning-based models, which may be susceptible to hallucinated signals due to distribution shifts in the training data.
    \item Optimized Signal-to-Noise Integration: Standard energy detectors with fixed bandwidths inevitably suffer from noise integration when the signal bandwidth is unknown, as they integrate noise over empty frequency bins. In contrast, our method dynamically estimates the interval of overlap $S$. By refining the start and end points of the signal, the detector maximizes the normalized average energy metric derived in Eq.~\ref{eq:Js_max}. This ensures that the detection statistic is computed primarily over signal-occupied bins, effectively maximizing the \ac{SNR} and resulting in higher detection probabilities compared to methods that cannot precisely localize the interval.
\end{itemize}

\section{Conclusion} \label{sec:conclusion}

In this paper, we investigate signal detection challenges in environments characterized by unknown signal bandwidth and occupancy intervals, particularly under adversarial and spectrum-sharing conditions. We introduced an effective \ac{GLRT}-based approach that leverages normalized average signal energy as a straightforward yet powerful detection metric. Our theoretical analysis provided bounds for false alarm and missed detection probabilities, explicitly linking them to the \ac{SNR} and the size of the signal set.
Addressing the significant computational overhead inherent in exhaustive search methods, we proposed a computationally efficient binary search algorithm that reduces the complexity from $O(N^2)$ to $O(N)$ in one-dimensional scenarios. Notably, this binary search method achieves performance comparable to exhaustive searches and demonstrates asymptotic consistency, ensuring that the interval-of-overlap converges to unity under constant \ac{SNR} conditions as the measurement size grows.
Our comprehensive simulation studies validated the efficacy and efficiency of the proposed binary search method. The method not only maintained near-optimal performance relative to exhaustive searches but also demonstrated superior detection capabilities and significantly lower computational complexity compared to contemporary neural network-based approaches, notably outperforming specialized U-Net models.
In conclusion, our proposed \ac{GLRT}-based binary search approach provides a robust and computationally efficient solution for signal detection in uncertain spectral environments, promising substantial practical benefits for spectrum sensing applications, especially in resource-constrained systems. Future work will include comparisons with benchmark receivers that exploit known structural attributes. Additionally, significant attention will be directed towards addressing the important case of multiple simultaneous signals and the development of more realistic models for signals that are not exactly aligned with the degrees of freedom.

\appendices
\section{Proof of Lemma~\ref{lem:likelihood}}
\label{sec:likelihood_proof} 
Given any signal region $S$ and \ac{SNR} $\gamma$,
each $X_n$ is exponentially distributed as
\begin{equation}
    p(X_n|S,\gamma) = \lambda_n^{-1} e^{-
    X_n/\lambda_n},
\end{equation}
where
\begin{equation}
    \lambda_n = \begin{cases}
        1 & \mbox{if } n \not \in S, \\
        1+\gamma & \mbox{if } n \in S, 
    \end{cases}
\end{equation}
Therefore, 
\begin{align}
    \MoveEqLeft \log \frac{p(\bs{X}|S,\gamma)}{p_0(\bs{X})} =  
    \log \prod_{n \in S} \left[ \frac{1}{1+\gamma}
    \exp(X_n(1-1/(1+\gamma)) \right] \nonumber \\
    &= -|S|\log(1 + \gamma) + 
    \left[ \frac{\gamma}{1+\gamma} \right] \sum_n X_n 
    \nonumber \\
    &= |S| \left[ \wb{X}_S\left[ \frac{\gamma}{1+\gamma} \right]  - \log(1+\gamma) \right], \nonumber
\end{align}
which proves \eqref{eq:Jsgam}.
To find the maxima over $\gamma$, we take the
derivative with respect to $\gamma$:
\begin{align}
    \MoveEqLeft\frac{1}{1+\gamma} = \wb{X}_S\frac{1}{(1+\gamma)^2}
    \nonumber \\
    & \Rightarrow  1+\gamma = \wb{X}_S.
\end{align}
This equation will have a solution $\gamma \geq 0$
if and only if $\wb{X}_S \geq 1$.
When $\wb{X}_S \geq 1$, the maximizing value
is $\gamma = \wb{X}_S-1$. Substituting
this expression into \eqref{eq:Jsgam}, we
obtain:
\begin{align}
    J(S) &= \max_{\gamma \geq 0} J(S,\gamma)
    \nonumber \\
    &=  |S| \left[ 
    \wb{X}_S -1 - \log(\wb{X}_S) \right]
    \mbox{ when } \wb{X}_S \geq 1.
     \label{eq:JS1}
\end{align}
When $\wb{X}_S \leq 1$, the maximum value
of $J(S,\gamma)$ occurs when $\gamma = 0$
, in which case:
\begin{equation} \label{eq:JS2}
    J(S) = 0 \mbox{ when } \wb{X}_S < 1.
\end{equation}
The formula \eqref{eq:Js_max} matches
\eqref{eq:JS1} when $\wb{X}_S\geq 1$
\eqref{eq:JS2} when $\wb{X}_S < 1$.

\section{Proof of Lemma~\ref{lem:pfa}}
\label{sec:pfa_proof}

The proof is a simple application
of a union bound:
\begin{align}
   \MoveEqLeft P_{\subsf FA} = \mathbb{P}\left( \wh{H}=1|H=0\right) \nonumber \\
    &=  \mathbb{P}\left( \wb{X}^+_S \geq u_{|S|} \mbox{ for some } S|H=0\right) \nonumber \\
   &\leq \sum_{S} \mathbb{P}\left( 
   \wb{X}^+_S \geq u_{|S|}|H=0\right) \label{eq:pfabnd1}
\end{align}
Hence, we can approximate
the bound \eqref{eq:pfabnd1} as:
\begin{align}
     P_{\subsf FA} \leq \frac{N^2}{2} \max_{S} \mathbb{P}\left( \wb{X}^+_S \geq u_{|S|}|H=0\right) \label{eq:pfabnd2}.
\end{align}
Also, from \eqref{eq:Xs}, we have that
\[
    |S|\wb{X}_S = \sum_{n \in S} X_n.
\]
Under the null hypothesis, $H=0$, the values $X_n$
are i.i.d. exponential random variables with 
a mean $\Exp(X_n)=1$.  Thus,
\[
    Z_S := 2|S|\wb{X}_S 
\]
is a chi-squared random variable with $2|S|$
degrees of freedom~\cite{leemis1995reliability}.
Therefore, for any set $S$ with $|S|=\ell$,
\begin{align}
  \MoveEqLeft \mathbb{P}\left( \wb{X}^+_S \geq u_{|S|}|H=0\right) = \mathbb{P}\left( \wb{X}^+_S \geq u_{\ell} |H=0\right) \nonumber \\
  &=  \mathbb{P}\left( \wb{X}_S \geq u_{\ell} |H=0\right) \nonumber \\
  &= \mathbb{P}\left( Z_S \geq 2\ell u_{\ell} |H=0\right)
  = F(2{\ell} u_{\ell}; 2\ell). \label{eq:pfabnd3}.
\end{align}
Substituting \eqref{eq:pfabnd3} into
\eqref{eq:pfabnd2}, we obtain \eqref{eq:pfabnd}.

\section{Proof of Lemma~\ref{lem:pmd}}
\label{sec:proof_pmd}
Let $S$ and $\gamma$ be the true signal region
and \ac{SNR} under the signal present hypothesis $H=1$.
The probability of missed detection is:
\begin{align}
     \MoveEqLeft P_{\subsf MD} = 
     \mathbb{P}(\wh{H}=0|H=1) \nonumber \\
     &= \mathbb{P}(\wb{X}_{S'} < u_{|S'|} \mbox{ for all } S') \nonumber \\
     &\leq \mathbb{P}(\wb{X}_{S} \leq u_{|S|}).
     \label{eq:pmdbnd1}
\end{align}
That is, we can bound the missed detection 
probability by the behavior on the true set.
Now, under the hypothesis $H=1$,
for all elements $n \in S$,
$X_n$ is exponentially distributed with
$\Exp(X_n) = (1+\gamma)$.  Hence,
\[
    Z_S = 2|S| \wb{X}_S / (1+\gamma)
\]
is chi-squared distributed with $2|S|$
degrees of freedom.  Therefore, using
the complementary \ac{CDF} in \eqref{eq:ccdf_cs},
we can write the probability bound in \eqref{eq:pmdbnd1}
as
\begin{equation}
    P_{\subsf MD} \leq 1 - F\left(\frac{2\ell u_{\ell}}{1+\gamma}; 2\ell \right),
\end{equation}
where $\ell=|S|$.

\section{Proof of Theorem \ref{thm:iou}}
\label{sec:proof_iou}

\subsection{Equivalent Algorithm}
Before proving the theorem, we first rewrite
Algorithm.~\ref{alg:int_estimation}
in the form of Algorithm.~\ref{alg:int_est_equiv} which is easier to analyze.
The following lemma shows that these two algorithms
are equivalent; hence, we can focus
on Algorithm.~\ref{alg:int_est_equiv}
for the subsequent analysis.

\begin{algorithm}
\caption{Equivalent Binary Interval Estimation}\label{alg:int_est_equiv}
\begin{algorithmic}[1]
\Require Interval length $N = 2^M$
\Require Dyadic powers $Z_{m,i}$
\State $\wh{a}^{(0)} \gets 0, 
\wh{b}^{(0)} \gets N$ 
\item $J_{\rm max} = J([\wh{a}^{(0)},\wh{b}^{(0)}))$
\State $\ell_0 \gets 2^M$ // initial sub-interval length

\For{$t=0,\ldots,M-1$}

    \State // Get layer info
    \State $m \gets M-t-1$ // layer    
    \State $\ell_{t+1} \gets 2^m$ // sub-interval length

    \State
    \State // Optimization over the left boundary
    \State $\wh{a}^{(t+1)} \gets \wh{a}^{(t)}$
    \For {$\delta \in \{-1, 1\}$}
        \State $a' \gets \wh{a}^{(t)} + \delta \ell_{t+1}$
        \State $S' = [a', \wh{b}^{(t)})$
        \State $J' \gets J(S')$
        \If {$J' > J_{\rm max}$, $|S'| > 0$, \textbf{and} $a'\geq 0$}
            \State $J_{\rm max} \gets J'$
            \State $\wh{a}^{(t+1)} \gets a'$
        \EndIf
    \EndFor

    \State
    \State // Optimization over the right boundary
    \State $\wh{b}^{(t+1)} \gets \wh{b}^{(t)}$
    \For {$\delta \in \{-1, 1\}$}
        \State $b' \gets \wh{b}^{(t)} + \delta \ell_{t+1}$
        \State $S' = [\wh{a}^{(t+1)}, b')$
        \State $J' \gets J(S')$
        \If {$J' > J_{\rm max}$, $|S'| > 0$, \textbf{and} $b'\leq N$}
            \State $J_{\rm max} \gets J'$
            \State $\wh{b}^{(t+1)} \gets b'$
        \EndIf
    \EndFor
\EndFor
\end{algorithmic}
\end{algorithm}

\begin{lemma} \label{lem:equiv}
Fix $N$ and consider any power measurements
$X_n$. Let 
\begin{equation} \label{eq:abbin}
    (\wh{a}^{(t)}_0,\wh{b}^{(t)}_0) = (\wh{a}^{(t)},\wh{b}^{(t)})
\end{equation} 
denote 
the estimates generated from 
Algorithm.~\ref{alg:int_estimation},
and let 
\begin{equation} \label{eq:abequiv}
    (\wh{a}^{(t)}_1,\wh{b}^{(t)}_1) = (\wh{a}^{(t)},\wh{b}^{(t)})
\end{equation} 
be the outputs of Algorithm.~\ref{alg:int_est_equiv}
with input $Z_{m,i}$ in \eqref{eq:Zmi}.
Then, for all $t=0,\ldots,M-1$:
\begin{equation} \label{eq:abequal}
    (\wh{a}^{(t)}_1,\wh{b}^{(t)}_1) = (\wh{a}^{(t)}_0,\wh{b}^{(t)}_0).
\end{equation} 
\end{lemma}
\begin{proof}

We use induction for the proof. For $t=0$, we have:
\begin{align}
\wh{a}_0^{(0)}=\wh{i}^{(0)} \ell_{0} = 0&, \quad \wh{b}_0^{(0)}=\wh{j}^{(0)} \ell_{0}=N, \notag \\
\wh{a}_1^{(0)}=0 &, \quad \wh{b}_1^{(0)}=N
\end{align}
Suppose that the argument holds for step $t$, so we have:
\begin{equation} \label{eq:abequal}
    (\wh{a}^{(t)}_1,\wh{b}^{(t)}_1) = (\wh{a}^{(t)}_0,\wh{b}^{(t)}_0).
\end{equation} 
we intend to prove that it will also hold for step $t+1$. We prove the equivalence of $\wh{a}^{(t+1)}_0$ and $\wh{a}^{(t+1)}_1$; for $\wh{b}$, it will be exactly the same.
\begin{align}
    \wh{a}^{(t+1)}_0 &= \wh{i}^{(t+1)} \ell_{(t+1)} = (2\wh{i}^{(t)}+\delta_{max}) \ell_{(t+1)} \notag \\
    \wh{a}^{(t+1)}_1 &= \wh{a}^{(t)}_1 + \delta_{max} \ell_{(t+1)} \notag \\
\end{align}
where $\delta_{max}$ is the $\delta$ that maximizes the objective function $J'$, which is exactly the same for both algorithms. But by assumption, we know that $\wh{a}^{(t)}_1 = \wh{a}^{(t)}_0$, and by Algorithm~\ref{alg:int_estimation}, we also know that $\wh{a}^{(t)}_0 = \wh{i}^{(t)} \ell_{t}$ and $\ell_{(t+1)}=2\ell_{t}$. So we'll have:
\begin{align}
    \wh{a}^{(t+1)}_1 &= \wh{a}^{(t)}_1 + \delta_{max} \ell_{(t+1)} = \wh{i}^{(t)} 2\ell_{(t+1)} + \delta_{max} \ell_{(t+1)} \notag \\
    &= (2\wh{i}^{(t)}+\delta_{max}) \ell_{(t+1)} = \wh{a}^{(t+1)}_0
\end{align}
which proves the equivalency of $\wh{\alpha}_0$ and $\wh{\alpha}_1$. The equivalency of $\wh{\beta}_0$ and $\wh{\beta}_1$ could be proved exactly with the same logic.
\end{proof}

\subsection{Asymptotic Likelihood Function}
In this subsection, we derive 
a simple expression for the likelihood function
in the limit of large $N$.
The likelihood function \eqref{eq:Js_max} implicitly depends on $N$.
To make this dependence explicit, we will write $J_N(S)$ for $J(S)$
where the set $S \subseteq \{0,1,\ldots,N-1\}$.
Next, for any $\alpha, \beta$ with
\begin{equation} \label{eq:abinterval}
    0 \leq \alpha \leq \beta \leq 1,
\end{equation}
we define the sequences 
\begin{equation} \label{eq:Snab1}
    a_N = \lfloor \alpha N \rfloor, \quad
    b_N = \lfloor \beta N \rfloor,
\end{equation}
and let $S_N$ denote the interval
\begin{equation} \label{eq:Snab2}
    S_N = [a_N, b_N).
\end{equation}
Note that, by the definition of \eqref{eq:Struelim}, $S^0_N$ is $S_N$
for $\alpha=\alpha^0$ and $\beta=\beta^0$.
Also, define the function:
\begin{equation} \label{eq:Gndef}
    G_N(\alpha,\beta) := \frac{1}{N} J_N(\lfloor \alpha N \rfloor, \lfloor \beta N \rfloor),
\end{equation}
which represents a normalized version of the likelihood function 
on the interval $S_N$.

\begin{lemma} \label{lem:Jslim}
For any $\alpha$ and $\beta$ satisfying \eqref{eq:abinterval} and
\begin{equation} \label{eq:aboverlap1}
    \alpha \leq \beta^0 \mbox{ and } \beta \geq \alpha^0, 
\end{equation}
we have that:
\begin{align}
    \lim_{N \rightarrow \infty} \wb{X}_{S_N} &= Z(\alpha, \beta), 
    \label{eq:Xlim} \\
    \lim_{N \rightarrow \infty} G_N(\alpha,\beta) &= G(\alpha, \beta)
    \label{eq:Jlim}
\end{align}
where the convergence is almost surely and
\begin{align}
    Z(\alpha,\beta) &:= 1 + \gamma
        \frac{\min\{\beta_0,\beta\} - \max\{\alpha_0,\alpha\}}{\beta-\alpha} 
        \label{eq:Zdeflim} \\
    G(\alpha,\beta) &:= (\beta-\alpha)
    \left[Z(\alpha,\beta) - 1 - \log Z(\alpha,\beta) \right]
    \label{eq:Gdeflim}
\end{align}
\end{lemma}
\begin{proof}
For any $S$, $\wb{X}_S$ in \eqref{eq:Xs} is the average of values $X_n$ with $n \in S$.
Therefore, 
the expected values in \eqref{eq:xscalar} show that
\begin{equation} \label{eq:expXs1}
    \Exp(\wb{X}_S) = 1 + \frac{|S \cap S^0_N|}{|S|}\gamma.
\end{equation}
From the definitions of $S^0_N$ in \eqref{eq:Struelim} and  $S_N$ in \eqref{eq:Snab1}
and \eqref{eq:Snab2}, as well as the condition \eqref{eq:aboverlap1},
we have that
\begin{subequations}
\begin{align}
    \lim_{N \rightarrow \infty} \frac{|S_N \cap S^0_N|}{N} &=  \min\{\beta_0,\beta\} - \max\{\alpha_0,\alpha\}\\
    \lim_{N \rightarrow \infty} \frac{|S_N|}{N} &= \beta-\alpha.
\end{align}
\end{subequations}
Also, since $\wb{X}_{S_N}$ is an average of
i.i.d. random variables, it follows from the
strong law of large numbers that
\begin{align}
   \MoveEqLeft \lim_{N \rightarrow \infty} \wb{X}_{S_N}
    = \lim_{N \rightarrow \infty} \Exp(\wb{X}_{S_N}) \nonumber \\
    &= 1 + \gamma  \lim_{N \rightarrow \infty}
    \frac{|S \cap S^0_N|}{|S|} \nonumber \\
    &= 1 + \gamma  \frac{\min\{\beta_0,\beta\} - \max\{\alpha_0,\alpha\}}{\beta-\alpha}
    \nonumber \\
    &= Z(\alpha,\beta),
\end{align}
which proves \eqref{eq:Xlim}.  
From (\ref{eq:Gndef}) and (\ref{eq:Js_max}) we have that:
\begin{align} \label{}
    G_N(\alpha,\beta) &= \frac{1}{N} J_N([\lfloor \alpha N \rfloor, \lfloor \beta N \rfloor) ) \notag \\
    &= \frac{\lfloor \beta N \rfloor - \lfloor \alpha N \rfloor}{N} \left[ \wb{X}_{S_N} -1- \log(\wb{X}_{S_N}) \right]
\end{align}
From (\ref{eq:Xlim}), this limit is given by:
\begin{align}
\MoveEqLeft \lim_{N \rightarrow \infty} G_N(\alpha,\beta) &= (\beta-\alpha)
    \left[Z(\alpha,\beta) - 1 - \log Z(\alpha,\beta) \right] \notag \\
    &= G(\alpha, \beta)
\end{align}
which proves (\ref{eq:Jlim}).
\end{proof}

We will call the function $G(\alpha,\beta)$
the \emph{asymptotic normalized likelihood}
and $Z(\alpha,\beta)$ the \emph{asymptotic
average signal energy}.  An important property
of the asymptotic normalized
likelihood function is that it possesses a certain \emph{triangular
maximization} property, as given by the following
definition.

\begin{definition} 
A scalar-valued function $f(x)$ of a scalar $x$
in some interval $A$ 
has a  \emph{triangular maxima} at $x=x^*$
with constant $c > 0$ 
if:
\begin{enumerate}[label=(\alph*)]    
    \item $f'(x) > c$ in the region $x < x^*$ and $x \in A$. 
    \item $f'(x) < -c$ in the region $x > x^*$ and $x \in A$
\end{enumerate}
\end{definition}

Note that if $x^*$ is a triangular maxima in an interval $A$, then
$x^* = \argmax f(x)$ for $x \in A$.

\begin{lemma} \label{lem:tri_max}  Fix any $\alpha_0 < \beta_0$ and $\gamma$, and consider the asymptotic normalized likelihood
function $G(\alpha,\beta)$ in \eqref{eq:Gdeflim}.
There exists a constant $c$, possibly
dependent on $\gamma$, such that:
\begin{enumerate}[label=(\alph*)]
    \item For fixed $\alpha \leq  \beta^0$, $G(\alpha,\beta)$
    has a triangular maximum at $\beta=\beta^0$
    with constant $c$ in the region $\beta \geq \alpha^0$.
     \item For fixed $\beta \geq \alpha^0$, $G(\alpha,\beta)$
    has a triangular maximum at $\alpha=\alpha^0$
    with constant $c$ in the region $\alpha \leq \beta^0$.
\end{enumerate}
\end{lemma}

\begin{proof}
We prove the first part with a fixed $\alpha$; the second part can be proved similarly. We can write $Z(\alpha, \beta)$ as:
\begin{equation} \label{eq:Z_beta}
    Z(\alpha, \beta) =
    \begin{cases} 
        1 + \gamma \frac{\beta_0 - \wh{\alpha}}{\beta-\alpha}, & \text{if } \beta \geq \beta_0 \\
        1 + \gamma \frac{\beta - \wh{\alpha}}{\beta-\alpha}, & \text{if } \beta < \beta_0
    \end{cases}
\end{equation}
where $\wh{\alpha} = \max\{\alpha, \alpha^0\}$. 
So the derivatives of $Z$ with respect to $\beta$ will be:
\begin{equation} \label{eq:Z_beta_partial}
    \frac{\partial Z}{\partial \beta} =
    \begin{cases}
        -\gamma \frac{\beta_0 - \wh{\alpha}}{(\beta-\alpha)^2}, & \text{if } \beta \geq \beta_0 \\
        \gamma \frac{\wh{\alpha} - \alpha}{(\beta-\alpha)^2}, & \text{if } \beta < \beta_0
    \end{cases}
\end{equation}
On the other hand, according to (\ref{eq:Gdeflim}), we can write the partial derivatives of $G$ as:
\begin{equation}
    \frac{\partial G}{\partial \beta} = Z(\beta) - 1 - \log Z(\beta) + (\beta-\alpha) \frac{\partial Z}{\partial \beta} \left[1-\frac{1}{Z(\beta)} \right]
\end{equation}
From (\ref{eq:Z_beta}) and (\ref{eq:Z_beta_partial}) we have:
\begin{equation}
    \frac{\partial G}{\partial \beta} = 
    \begin{cases} 
        f_1(Z), & \text{if } \beta \geq \beta_0 \\
        f_2(Z,\beta) , & \text{if } \beta < \beta_0
    \end{cases}
\end{equation}
where:
\begin{align}
    f_1(Z) &= -\log(Z)+1-\frac{1}{Z} \notag \\
    f_2(Z,\beta) &= Z-1-\log(Z)+\gamma\frac{\wh{\alpha}-\alpha}{\beta-\alpha}\left[ 1-\frac{1}{Z} \right]
\end{align}
We can rewrite $f_1(Z)$ as:
\begin{equation}
f_1(Z) = \log\left(\frac{1}{Z}\right)+1-\frac{1}{Z} = -\left[ \frac{1}{Z} - 1 - \log(\left(\frac{1}{Z}\right) \right]
\end{equation}
From \eqref{eq:Gdeflim}, we know that $Z > 1 + \epsilon$ for some $\epsilon > 0$, and all $\beta$.  We also know that $s(t)=t-1-\log(t)$ is always positive and only zero at $t=1$. Hence, there is a constant $c > 0$ such that $f_1(Z) < -c$ for all $z \geq 1 + \epsilon$.  So,
\begin{equation}
\frac{\partial G}{\partial \beta}< -c \text{  for  } \beta > \beta_0.
\end{equation}
For the region $\beta \in [\alpha^0,\beta^0)$, we can break 
$f_2(Z,\beta)$ into two parts: $f_2^1(Z)$ and $f_2^2(Z,\beta)$, where:
\begin{align}
f_2^1(Z) &= Z-1-\log(Z)  \\
f_2^2(Z,\beta) &= \gamma\frac{\wh{\alpha}-\alpha}{\beta-\alpha}\left[ 1-\frac{1}{Z} \right].
\end{align}
We have $f_2^2(Z,\beta) \geq 0$ since $\wh{\alpha}=\max\{\alpha_0, \alpha\}$ and $Z>1$.  Also, similarly to above, we can show that $f_2^1(Z) > c$ for some $c$ and all $Z=Z(\alpha,\beta)$ with $\beta \in [\alpha^0,\beta^0]$.
Therefore, 
\begin{equation}
\frac{\partial G}{\partial \beta}>c  \text{  for  } \beta < \beta_0.
\end{equation}

\end{proof}

\subsection{Proof of Theorem~\ref{thm:iou}}
To make the dependence on $N$ explicit, let $\wh{a}^{(t)}_N$ and $\wh{b}^{(t)}_N$ be the outputs of the equivalent algorithm,
 Algorithm~\ref{alg:int_est_equiv}, with the input size $N$.
Let $\wh{\alpha}^{(t)}_N$ and $\wh{\beta}^{(t)}_N$
be the normalized values
\begin{equation} \label{eq:abndef}
    \wh{\alpha}^{(t)}_N = \frac{\wh{a}^{(t)}_N}{N}, \quad 
    \wh{\beta}^{(t)}_N = \frac{\wh{b}^{(t)}_N}{N}.
\end{equation}
Since $\wh{a}^{(t)}_N$ and $\wh{b}^{(t)}_N$ are integer multiples of
$\ell_t = N2^{-t}$, the normalized estimates 
$\wh{\alpha}^{(t)}$ and $\wh{\beta}^{(t)}$ will be at the discrete points:
\begin{equation} \label{eq:abdiscrete}
    \wh{\alpha}^{(t)} = \wh{i}^{(t)}2^{-t}, \quad 
    \wh{\beta}^{(t)} = \wh{j}^{(t)}2^{-t},    
\end{equation}
for some integer indices 
\begin{equation} \label{eq:ijdiscrete}
    \wh{i}^{(t)},~\wh{j}^{(t)} =0,\ldots 2^t.
\end{equation}
Now fix any $t_0 \geq 0$.
Since there are at most a finite number of choices in 
\eqref{eq:ijdiscrete}, Lemma~\ref{lem:Jslim} shows that,
with probability one, there exists an $N_0$ such that
\begin{equation} \label{eq:Gnapprox}
    |G_N( i2^{-t}, j2^{-t}) - G( i2^{-t}, j2^{-t})| < c 2^{-t-2},
\end{equation}
where $c$ is the constant in Lemma~\ref{lem:tri_max}, and all 
the inequalities are valid for all $i,j=0,\ldots,2^t$, $N \geq N_0$
, and $t \leq t_0$.  We are now ready to prove our main induction step.

\begin{lemma} \label{lem:ablim} Let $t_0$ and $N_0$ be defined as above.
With probability one, for all $N \geq N_0$ and $t \leq t_0$,
$\wh{\alpha}^{(t)}_N$ and $\wh{\beta}^{(t)}_N$ have a bounded distance
from the true values $\alpha^0$ and $\beta^0$ in the sense that:
 \begin{align}  \label{eq:ablim}
    |\wh{\alpha}^{(t)}_N - \alpha^0| \leq \delta_t, \quad
   |\wh{\beta}^{(t)}_N - \beta^0| \leq \delta_t 
\end{align}
where $\delta_t = 2^{-t}$.
\end{lemma}

\begin{proof}
We use induction for the proof. For $t=0$ we will have:
\begin{align}
    \big| \wh{\alpha}^{(0)} - \alpha_0 \big| \leq 1, \quad
    \big| \wh{\beta}^{(0)} - \beta_0 \big| \leq 1
\end{align}
since all values are in the interval $[0,1]$ 
and thus their distance cannot be larger than 1.
Suppose \eqref{eq:ablim} holds for the $t$-th step with $t < t_0$.
We prove that \eqref{eq:ablim} also holds for step $t+1$. 
The optimization over the left boundary in Algorithm.~\ref{alg:int_est_equiv}
is equivalent to 
\begin{align}
    \MoveEqLeft \wh{a}^{(t+1)}_N = \argmax_a J_N([a,\wh{b}^{(t)}_N)), \\
    & \mbox{s.t. } a = \wh{a}^{(t)}_N + \{0,\pm \ell_{t+1}\}.
\end{align}
Since $\ell_{t+1}=\delta_{t+1} N$, \eqref{eq:abndef} and \eqref{eq:Gndef} show that the maximization can be re-written as:
\begin{align} \label{eq:argmax_t_1}
    \MoveEqLeft \wh{\alpha}^{(t+1)}_N = \argmax_{\alpha} G_{N}(\alpha, \wh{\beta}^{(t)}_N) \nonumber \\
    & \text{s.t. } \wh{\alpha}^{(t+1)}_N = \wh{\alpha}^{(t)}_N
    + \{0,\pm \delta_{t+1}\}.
\end{align}
Now define
\begin{align} \label{eq:argmax_t_2}
    \MoveEqLeft \alpha^{(t+1)}_N = \argmax_{\alpha} G(\alpha, \wh{\beta}^{(t)}_N) \nonumber \\
    & \text{s.t. } \alpha^{(t+1)}_N = \wh{\alpha}^{(t)}_N
    + \{0,\pm \delta_{t+1}\},
\end{align}
which is identical to $\wh{\alpha}^{(t+1)}$, except that we have replaced
$G_N(\cdot)$ in the objective function with $G(\cdot)$, its asymptotic limit.
So, there are three possibilities for $\alpha^{(t+1)}_N$.
We will show that 
\begin{equation} \label{eq:abnd}
    |\wh{\alpha}^{(t+1)}_N - \alpha^0| \leq \delta_{t+1}
\end{equation}
for all three cases.

\noindent \textbf{Case 1:} $\alpha^{(t+1)}_N = \wh{\alpha}^{(t)}_N$.
In this case, we know that
\begin{subequations}
\begin{align}
G(\wh{\alpha}^{(t)}_N, \wh{\beta}^{(t)}_N) &\geq 
    G(\wh{\alpha}^{(t)}_N-\delta_{t+1}, \wh{\beta}^{(t)}_N) \\
G(\wh{\alpha}^{(t)}_N, \wh{\beta}^{(t)}_N) &\geq 
    G(\wh{\alpha}^{(t)}_N+\delta_{t+1}, \wh{\beta}^{(t)}_N) 
\end{align}
\end{subequations}
Since $G(\alpha,\wh{\beta}^{(t)}_N)$ has a triangular maximum
at $\alpha=\alpha^0$ it must be that $\alpha^0$ is in the interval
\begin{equation}
    \alpha^0  \in [\wh{\alpha}^{(t)}_N-\delta_{t+1} \wh{\alpha}^{(t)}_N+\delta_{t+1}].
\end{equation}
Hence either:
\begin{equation} \label{eq:aint1}
    \alpha^0  \in [\wh{\alpha}^{(t)}_N-\delta_{t+1}, \wh{\alpha}^{(t)}_N] 
\end{equation}
or 
\begin{equation} \label{eq:aint2}
    \alpha^0  \in [\wh{\alpha}^{(t)}_N, \wh{\alpha}^{(t)}_N+\delta_{t+1}].
\end{equation}
WLOG assume $\alpha^0$ is in the interval \eqref{eq:aint1}.
Therefore,
\begin{align}
   \MoveEqLeft G_N(\wh{\alpha}^{(t+1)}_N+\delta_{t+1}, \wh{\beta}^{(t)}_N) 
   \nonumber \\
   &\stackrel{(a)}{<}
    G(\wh{\alpha}^{(t+1)}_N+\delta_{t+1}, \wh{\beta}^{(t)}_N) + \frac{c}{2}\delta_{t+1} \nonumber \\
    &\stackrel{(b)}{\leq} G(\wh{\alpha}^{(t+1)}_N, \wh{\beta}^{(t)}_N) + \frac{c}{2}\delta_{t+1}
    - c\delta_{t+1} \nonumber \\
    &\stackrel{(c)}{<} G_N(\wh{\alpha}^{(t+1)}_N, \wh{\beta}^{(t)}_N) - \frac{c}{2}\delta_{t+1}
    +\frac{c}{2}\delta_{t+1} \nonumber \\
    &\leq G_N(\wh{\alpha}^{(t+1)}_N,    \wh{\beta}^{(t)}_N)  \label{eq:Gnmax1}
\end{align}
where (a) is due to \eqref{eq:Gnapprox};
(b) is due the fact that $G(\alpha,\beta)$ has a triangular maxima
at $\alpha^0 \leq \wh{\alpha}^{(t)}_N$; and 
(c) again is due to \eqref{eq:Gnapprox}.
Hence in the maximization \eqref{eq:argmax_t_1}, we have
that either $\wh{\alpha}^{(t+1)}$ satisfies
\begin{equation}
    \wh{\alpha}^{(t+1)} \in \{ \wh{\alpha}^{(t)}-\delta_{t+1}, \wh{\alpha}^{(t)} \}.
\end{equation}
Since by assumption $\alpha^0$ is in the interval
\eqref{eq:aint1}, it follows that \eqref{eq:abnd} holds.

\noindent \textbf{Case 2:}  $\alpha^{(t+1)}_N = \wh{\alpha}^{(t)}_N + \delta_{t+1}$.
In this case, we know that
\begin{subequations}
\begin{align}
G(\wh{\alpha}^{(t)}_N+\delta_{t+1}, \wh{\beta}^{(t)}_N) &\geq 
    G(\wh{\alpha}^{(t)}_N-\delta_{t+1}, \wh{\beta}^{(t)}_N) \\
G(\wh{\alpha}^{(t)}_N+\delta_{t+1}, \wh{\beta}^{(t)}_N) &\geq 
    G(\wh{\alpha}^{(t)}_N, \wh{\beta}^{(t)}_N).
\end{align}    
\end{subequations}
Since $G(\alpha,\wh{\beta}^{(t)}_N)$ has a triangular maximum
at $\alpha=\alpha^0$, it must be $\alpha^0 \geq \wh{\alpha}^{(t)}_N$.
Also, by the induction hypothesis, $\alpha^0 \leq \wh{\alpha}^{(t)}_N-\delta_t$.  Hence, $\alpha^0$ is in the interval:
\begin{equation}
    \alpha^0  \in [\wh{\alpha}^{(t)}_N, \wh{\alpha}^{(t)}_N+\delta_{t}].
\end{equation}
In particular, since $\delta_{t+1} = \delta_t / 2$, either:
\begin{equation} \label{eq:aint3}
    \alpha^0  \in [\wh{\alpha}^{(t)}_N, \wh{\alpha}^{(t)}_N+\delta_{t+1}] 
\end{equation}
or 
\begin{equation} \label{eq:aint4}
    \alpha^0  \in [\wh{\alpha}^{(t)}_N+\delta_{t+1}, \wh{\alpha}^{(t)}_N+2\delta_{t+1}].
\end{equation}
Suppose that $\alpha^0$ is in the interval \eqref{eq:aint3}.
Then, a similar argument as \eqref{eq:Gnmax1} shows that
\begin{equation} \label{eq:Gnmax2}
   G_N(\wh{\alpha}^{(t+1)}_N-\delta_{t+1}, \wh{\beta}^{(t)}_N) 
    \leq G_N(\wh{\alpha}^{(t+1)}_N, \wh{\beta}^{(t)}_N)  
\end{equation}
which shows that$\wh{\alpha}^{(t)}_N$ must be one of the two values:
\begin{equation}
    \wh{\alpha}^{(t+1)} \in \{ \wh{\alpha}^{(t)}, \wh{\alpha}^{(t)} 
    + \delta_{t+1}\}.
\end{equation}
Since $\alpha^0$ is the interval \eqref{eq:aint3} we see that \eqref{eq:abnd}
is satisfied.  Similarly, if $\alpha^0$ is in the interval \eqref{eq:aint4}.
one can show that
\begin{equation} \label{eq:Gnmax3}
   G_N(\wh{\alpha}^{(t+1)}_N, \wh{\beta}^{(t)}_N) 
    \leq G_N(\wh{\alpha}^{(t+1)}_N+\delta_{t+1}, \wh{\beta}^{(t)}_N),
\end{equation}
which shows that $\wh{\alpha}^{(t)}_N$ must be one of the two values:
\begin{equation}
    \wh{\alpha}^{(t+1)} \in \{ \wh{\alpha}^{(t)}+\delta_{t+1}, \wh{\alpha}^{(t)} 
    + 2\delta_{t+1}\}.
\end{equation}
Since $\alpha^0$ is the interval \eqref{eq:aint4} we see that \eqref{eq:abnd}
is satisfied. 

\noindent \textbf{Case 3:}  $\alpha^{(t+1)}_N = \wh{\alpha}^{(t)}_N + \delta_{t+1}$. Similarly to case 2, we can show that \eqref{eq:abnd}
is satisfied.

Hence, we have shown that, in all three cases \eqref{eq:abnd}
is satisfied.  The proof for $|\wh{\beta}^{(t+1)}_N -\beta^0| \leq \delta_{t+1}$ is similar.  So, by induction, \eqref{eq:ablim} is satisfied
for all $N \geq N_0$ and $t \leq t_0$ with probability one.
\end{proof}

Now we proceed to prove (\ref{eq:ioubnd}). We have:
\begin{align}
    \mathrm{IoU}^{(t)}_N &= \frac{|\wh{S}^{(t)}_N \cap S^0_N|}{|\wh{S}^{(t)}_N \cup S^0_N|}
\end{align}
On the other hand, we have:
\begin{align}
    S^0_N &= [a^0_N, b^0_N), \; a^0_N = \lfloor{\alpha_0 N} \rfloor, \; 
    b^0_N = \lfloor{\beta_0 N} \rfloor, \notag \\
    \wh{S}^{(t)}_N &= [\wh{a}^{(t)}_N, \wh{b}^{(t)}_N), \; a^{(t)}_N = \wh{\alpha}^{(t)}_N N, \; b^{(t)}_N = \wh{\beta}^{(t)}_N N
\end{align}
So, at the limit, we'll have:
\begin{align} \label{eq:iou1}
    \lim_{N \rightarrow \infty} \mathrm{IoU}^{(t)}_N = \lim_{N \rightarrow \infty} \frac{\big|\min(\wh{\beta}^{(t)}_N,\beta_0) - \max(\wh{\alpha}^{(t)}_N, \alpha_0)\big|}{\big|\max(\wh{\beta}^{(t)}_N,\beta_0) - \min(\wh{\alpha}^{(t)_N}, \alpha_0)\big|}
\end{align}
Lemma~\ref{lem:ablim} and the triangle inequality show:
\begin{align}
    & \lim_{N \rightarrow \infty} \min|\{\wh{\beta}^{(t)},\beta_0\}  \geq \beta^0 - \big|\wh{\beta}^{(t)} - \beta^0 \big| \geq 
    \beta^0-2^{-t} \notag \\
    & \lim_{N \rightarrow \infty} \max|\{\wh{\beta}^{(t)},\beta_0\} \leq \beta^0 + \big| \wh{\beta}^{(t)} - \beta_0 \big|
    \leq \beta^0 - 2^{-t}\notag \\
    & \lim_{N \rightarrow \infty} \min|\{\wh{\alpha}^{(t)},\alpha\} \geq \beta^0 - \big|\wh{\alpha}^{(t)} - \alpha^0 \big| \geq 
    \beta^0-2^{-t} \notag \\
    & \lim_{N \rightarrow \infty} \max|\{\wh{\alpha}^{(t)},\alpha\} \leq \alpha^0 + \big| \wh{\alpha}^{(t)} - \alpha \big|
    \leq \alpha^0 - 2^{-t}.\notag
\end{align}
Since the nominator and denominator of the $IoU$ in \eqref{eq:iou1} are positive, we can bound its limit as follows:
\begin{align}
\MoveEqLeft \lim_{N \rightarrow \infty} \mathrm{IoU}^{(t)}_N \geq \frac{\big|\beta_0 - \alpha_0\big| - 2^{(-t+1)}}{\big|\beta_0 - \alpha_0\big| + 2^{(-t+1)}} \notag \\
&= 1-\frac{2^{(-t+2)}}{\big|\beta_0 - \alpha_0\big| + 2^{(-t+1)}}
\geq 1- \frac{2^{(-t+2)}}{\beta_0 - \alpha_0},
\end{align}
where the limit is true almost surely for any fixed $t$.
Since there are a countable number of values of $t$, it follows that
\begin{align}
    \lim_{N,t \rightarrow \infty} \mathrm{IoU}^{(t)}_N=1
\end{align}
almost surely, which proves (\ref{thm:iou}).

\bibliographystyle{IEEEtran}
\bibliography{refs/bibliography}{}

\begin{thebibliography}{10}
\providecommand{\url}[1]{#1}
\csname url@samestyle\endcsname
\providecommand{\newblock}{\relax}
\providecommand{\bibinfo}[2]{#2}
\providecommand{\BIBentrySTDinterwordspacing}{\spaceskip=0pt\relax}
\providecommand{\BIBentryALTinterwordstretchfactor}{4}
\providecommand{\BIBentryALTinterwordspacing}{\spaceskip=\fontdimen2\font plus
\BIBentryALTinterwordstretchfactor\fontdimen3\font minus \fontdimen4\font\relax}
\providecommand{\BIBforeignlanguage}[2]{{%
\expandafter\ifx\csname l@#1\endcsname\relax
\typeout{** WARNING: IEEEtran.bst: No hyphenation pattern has been}%
\typeout{** loaded for the language `#1'. Using the pattern for}%
\typeout{** the default language instead.}%
\else
\language=\csname l@#1\endcsname
\fi
#2}}
\providecommand{\BIBdecl}{\relax}
\BIBdecl

\bibitem{ali2016advances}
A.~Ali and W.~Hamouda, ``Advances on spectrum sensing for cognitive radio networks: Theory and applications,'' \emph{IEEE communications surveys \& tutorials}, vol.~19, no.~2, pp. 1277--1304, 2016.

\bibitem{yucek2009survey}
T.~Yucek and H.~Arslan, ``A survey of spectrum sensing algorithms for cognitive radio applications,'' \emph{IEEE communications surveys \& tutorials}, vol.~11, no.~1, pp. 116--130, 2009.

\bibitem{nasser2021spectrum}
A.~Nasser, H.~Al~Haj~Hassan, J.~Abou~Chaaya, A.~Mansour, and K.-C. Yao, ``Spectrum sensing for cognitive radio: Recent advances and future challenge,'' \emph{Sensors}, vol.~21, no.~7, p. 2408, 2021.

\bibitem{janu2022machine}
D.~Janu, K.~Singh, and S.~Kumar, ``Machine learning for cooperative spectrum sensing and sharing: A survey,'' \emph{Transactions on Emerging Telecommunications Technologies}, vol.~33, no.~1, p. e4352, 2022.

\bibitem{lee2020detection}
K.-G. Lee and S.-J. Oh, ``Detection of frequency-hopping signals with deep learning,'' \emph{IEEE Communications Letters}, vol.~24, no.~5, pp. 1042--1046, 2020.

\bibitem{lin2022multi}
M.~Lin, X.~Zhang, Y.~Tian, and Y.~Huang, ``Multi-signal detection framework: A deep learning based carrier frequency and bandwidth estimation,'' \emph{Sensors}, vol.~22, no.~10, p. 3909, 2022.

\bibitem{gkelias2022gan}
A.~Gkelias and K.~K. Leung, ``Gan-based detection of adversarial em signal waveforms,'' in \emph{MILCOM 2022-2022 IEEE Military Communications Conference (MILCOM)}.\hskip 1em plus 0.5em minus 0.4em\relax IEEE, 2022, pp. 356--361.

\bibitem{kang2024cellular}
S.~Kang, M.~Mezzavilla, S.~Rangan, A.~Madanayake, S.~B. Venkatakrishnan, G.~Hellbourg, M.~Ghosh, H.~Rahmani, and A.~Dhananjay, ``Cellular wireless networks in the upper mid-band,'' \emph{IEEE Open Journal of the Communications Society}, 2024.

\bibitem{kang2024terrestrial}
S.~Kang, G.~Geraci, M.~Mezzavilla, and S.~Rangan, ``{Terrestrial-satellite spectrum sharing in the upper mid-band with interference nulling},'' in \emph{ICC 2024-IEEE International Conference on Communications}.\hskip 1em plus 0.5em minus 0.4em\relax IEEE, 2024, pp. 5057--5062.

\bibitem{soltanmohammadi2013spectrum}
E.~Soltanmohammadi, M.~Orooji, and M.~Naraghi-Pour, ``Spectrum sensing over mimo channels using generalized likelihood ratio tests,'' \emph{IEEE Signal Processing Letters}, vol.~20, no.~5, pp. 439--442, 2013.

\bibitem{li2018kernelized}
L.~Li, S.~Hou, and A.~L. Anderson, ``Kernelized generalized likelihood ratio test for spectrum sensing in cognitive radio,'' \emph{IEEE Transactions on Vehicular Technology}, vol.~67, no.~8, pp. 6761--6773, 2018.

\bibitem{prasad2020downscaled}
K.~S.~V. Prasad, K.~B. D’souza, and V.~K. Bhargava, ``A downscaled faster-{RCNN} framework for signal detection and time-frequency localization in wideband {RF} systems,'' \emph{IEEE Transactions on Wireless Communications}, vol.~19, no.~7, pp. 4847--4862, 2020.

\bibitem{tao2021radio}
M.~Tao, S.~Tang, J.~Li, X.~Zhang, Y.~Fan, and J.~Su, ``Radio frequency interference detection for sar data using spectrogram-based semantic network,'' in \emph{2021 IEEE International Geoscience and Remote Sensing Symposium IGARSS}.\hskip 1em plus 0.5em minus 0.4em\relax IEEE, 2021, pp. 1662--1665.

\bibitem{vagollari2021joint}
A.~Vagollari, V.~Schram, W.~Wicke, M.~Hirschbeck, and W.~Gerstacker, ``Joint detection and classification of rf signals using deep learning,'' in \emph{2021 IEEE 93rd Vehicular Technology Conference (VTC2021-Spring)}.\hskip 1em plus 0.5em minus 0.4em\relax IEEE, 2021, pp. 1--7.

\bibitem{li2022new}
W.~Li, K.~Wang, L.~You, and Z.~Huang, ``A new deep learning framework for hf signal detection in wideband spectrogram,'' \emph{IEEE Signal Processing Letters}, vol.~29, pp. 1342--1346, 2022.

\bibitem{benazzouza2024novel}
S.~Benazzouza, M.~Ridouani, F.~Salahdine, and A.~Hayar, ``A novel spectrogram based lightweight deep learning for iot spectrum monitoring,'' \emph{Physical Communication}, vol.~64, p. 102364, 2024.

\bibitem{zhang2024prompting}
Z.~Zhang, J.~An, N.~Ye, D.~Niyato, and K.~Yang, ``Prompting and tuning: In-band interference segmentation using segment anything model,'' \emph{IEEE Wireless Communications Letters}, 2024.

\bibitem{urkowitz2005energy}
H.~Urkowitz, ``Energy detection of unknown deterministic signals,'' \emph{Proceedings of the IEEE}, vol.~55, no.~4, pp. 523--531, 2005.

\bibitem{margoosian2015accurate}
A.~Margoosian, J.~Abouei, and K.~N. Plataniotis, ``An accurate kernelized energy detection in gaussian and non-gaussian/impulsive noises,'' \emph{IEEE Transactions on Signal Processing}, vol.~63, no.~21, pp. 5621--5636, 2015.

\bibitem{sobron2015energy}
I.~Sobron, P.~S. Diniz, W.~A. Martins, and M.~Velez, ``Energy detection technique for adaptive spectrum sensing,'' \emph{IEEE Transactions on Communications}, vol.~63, no.~3, pp. 617--627, 2015.

\bibitem{chatziantoniou2015threshold}
E.~Chatziantoniou, B.~Allen, and V.~Velisavljevic, ``Threshold optimization for energy detection-based spectrum sensing over hyper-rayleigh fading channels,'' \emph{IEEE Communications Letters}, vol.~19, no.~6, pp. 1077--1080, 2015.

\bibitem{umar2013unveiling}
R.~Umar, A.~U. Sheikh, and M.~Deriche, ``Unveiling the hidden assumptions of energy detector based spectrum sensing for cognitive radios,'' \emph{IEEE communications surveys \& tutorials}, vol.~16, no.~2, pp. 713--728, 2013.

\bibitem{ma2012matched}
L.~Ma, Y.~Li, and A.~Demir, ``Matched filtering assisted energy detection for sensing weak primary user signals,'' in \emph{2012 IEEE International Conference on Acoustics, Speech and Signal Processing (ICASSP)}.\hskip 1em plus 0.5em minus 0.4em\relax IEEE, 2012, pp. 3149--3152.

\bibitem{zhang2014matched}
X.~Zhang, R.~Chai, and F.~Gao, ``Matched filter based spectrum sensing and power level detection for cognitive radio network,'' in \emph{2014 IEEE global conference on signal and information processing (GlobalSIP)}.\hskip 1em plus 0.5em minus 0.4em\relax IEEE, 2014, pp. 1267--1270.

\bibitem{chaudhari2009autocorrelation}
S.~Chaudhari, V.~Koivunen, and H.~V. Poor, ``Autocorrelation-based decentralized sequential detection of ofdm signals in cognitive radios,'' \emph{IEEE Transactions on Signal Processing}, vol.~57, no.~7, pp. 2690--2700, 2009.

\bibitem{huang2013cyclostationarity}
G.~Huang and J.~K. Tugnait, ``On cyclostationarity based spectrum sensing under uncertain gaussian noise,'' \emph{IEEE Transactions on Signal Processing}, vol.~61, no.~8, pp. 2042--2054, 2013.

\bibitem{rebeiz2013optimizing}
E.~Rebeiz, P.~Urriza, and D.~Cabric, ``Optimizing wideband cyclostationary spectrum sensing under receiver impairments,'' \emph{IEEE Transactions on signal processing}, vol.~61, no.~15, pp. 3931--3943, 2013.

\bibitem{zeng2009eigenvalue}
Y.~Zeng and Y.-C. Liang, ``Eigenvalue-based spectrum sensing algorithms for cognitive radio,'' \emph{IEEE transactions on communications}, vol.~57, no.~6, pp. 1784--1793, 2009.

\bibitem{quan2008optimal}
Z.~Quan, S.~Cui, A.~H. Sayed, and H.~V. Poor, ``Optimal multiband joint detection for spectrum sensing in cognitive radio networks,'' \emph{IEEE transactions on signal processing}, vol.~57, no.~3, pp. 1128--1140, 2008.

\bibitem{farhang2008filter}
B.~Farhang-Boroujeny, ``Filter bank spectrum sensing for cognitive radios,'' \emph{IEEE Transactions on signal processing}, vol.~56, no.~5, pp. 1801--1811, 2008.

\bibitem{lin2010progressive}
M.~Lin, A.~P. Vinod, and C.-M. Samson, ``Progressive decimation filter banks for variable resolution spectrum sensing in cognitive radios,'' in \emph{2010 17th International Conference on Telecommunications}.\hskip 1em plus 0.5em minus 0.4em\relax IEEE, 2010, pp. 857--863.

\bibitem{el2013improved}
S.~E. El-Khamy, M.~S. El-Mahallawy, and E.-N.~S. Youssef, ``Improved wideband spectrum sensing techniques using wavelet-based edge detection for cognitive radio,'' in \emph{2013 international conference on computing, networking and communications (ICNC)}.\hskip 1em plus 0.5em minus 0.4em\relax IEEE, 2013, pp. 418--423.

\bibitem{jindal2014wavelet}
S.~Jindal, D.~Dass, and R.~Gangopadhyay, ``Wavelet based spectrum sensing in a multipath rayleigh fading channel,'' in \emph{2014 Twentieth national conference on communications (NCC)}.\hskip 1em plus 0.5em minus 0.4em\relax IEEE, 2014, pp. 1--6.

\bibitem{martone2002multiresolution}
M.~Martone, ``Multiresolution sequence detection in rapidly fading channels based on focused wavelet decompositions,'' \emph{IEEE Transactions on Communications}, vol.~49, no.~8, pp. 1388--1401, 2002.

\bibitem{mohammadi2024parallel}
Z.~Mohammadi and A.~Zaimbashi, ``Parallel multiband spectrum sensing in lte-based cognitive radios,'' \emph{IEEE Transactions on Vehicular Technology}, vol.~73, no.~9, pp. 13\,193--13\,205, 2024.

\bibitem{kumar2022intelligent}
A.~Kumar and H.~Sharma, ``Intelligent cognitive radio spectrum sensing based on energy detection for advanced waveforms,'' \emph{Radioelectronics and Communications Systems}, vol.~65, no.~3, pp. 149--154, 2022.

\bibitem{vluadeanu2024average}
C.~Vl{\u{a}}deanu, O.~M.~K. Al-Dulaimi, A.~Mar{\c{t}}ian, and D.~C. Popescu, ``Average energy detection with adaptive threshold for spectrum sensing in cognitive radio systems,'' \emph{IEEE Transactions on Vehicular Technology}, 2024.

\bibitem{tian2006wavelet}
Z.~Tian and G.~B. Giannakis, ``A wavelet approach to wideband spectrum sensing for cognitive radios,'' in \emph{2006 1st international conference on cognitive radio oriented wireless networks and communications}.\hskip 1em plus 0.5em minus 0.4em\relax IEEE, 2006, pp. 1--5.

\bibitem{mallat2002singularity}
S.~Mallat and W.~L. Hwang, ``Singularity detection and processing with wavelets,'' \emph{IEEE transactions on information theory}, vol.~38, no.~2, pp. 617--643, 2002.

\bibitem{stephane1999wavelet}
M.~Stephane, ``A wavelet tour of signal processing,'' 1999.

\bibitem{wang2023comparisons}
Y.~Wang and P.~He, ``Comparisons between fast algorithms for the continuous wavelet transform and applications in cosmology: The 1d case,'' \emph{RAS Techniques and Instruments}, vol.~2, no.~1, pp. 307--323, 2023.

\bibitem{ieng2018complexity}
S.-H. Ieng, E.~Lehtonen, and R.~Benosman, ``Complexity analysis of iterative basis transformations applied to event-based signals,'' \emph{Frontiers in neuroscience}, vol.~12, p. 373, 2018.

\bibitem{kumar2016improved}
A.~Kumar, S.~Saha, and R.~Bhattacharya, ``Improved wavelet transform based edge detection for wide band spectrum sensing in cognitive radio,'' in \emph{2016 USNC-URSI Radio Science Meeting}.\hskip 1em plus 0.5em minus 0.4em\relax IEEE, 2016, pp. 21--22.

\bibitem{kobayashimultiband}
R.~T. Kobayashi, L.~Claudino, A.~G. Hernandes, and T.~Abr{\~a}o, ``Multiband spectrum sensing via edge detection using a wavelet approach.''

\bibitem{tian2007compressed}
Z.~Tian and G.~B. Giannakis, ``Compressed sensing for wideband cognitive radios,'' in \emph{2007 IEEE International conference on acoustics, speech and signal processing-ICASSP'07}, vol.~4.\hskip 1em plus 0.5em minus 0.4em\relax IEEE, 2007, pp. IV--1357.

\bibitem{tropp2007signal}
J.~A. Tropp and A.~C. Gilbert, ``Signal recovery from random measurements via orthogonal matching pursuit,'' \emph{IEEE Transactions on information theory}, vol.~53, no.~12, pp. 4655--4666, 2007.

\bibitem{dai2009subspace}
W.~Dai and O.~Milenkovic, ``Subspace pursuit for compressive sensing signal reconstruction,'' \emph{IEEE transactions on Information Theory}, vol.~55, no.~5, pp. 2230--2249, 2009.

\bibitem{venkataramani2000perfect}
R.~Venkataramani and Y.~Bresler, ``Perfect reconstruction formulas and bounds on aliasing error in sub-nyquist nonuniform sampling of multiband signals,'' \emph{IEEE Transactions on Information Theory}, vol.~46, no.~6, pp. 2173--2183, 2000.

\bibitem{cabric2006spectrum}
D.~Cabric, A.~Tkachenko, and R.~W. Brodersen, ``Spectrum sensing measurements of pilot, energy, and collaborative detection,'' in \emph{Milcom 2006-2006 IEEE military communications conference}.\hskip 1em plus 0.5em minus 0.4em\relax IEEE, 2006, pp. 1--7.

\bibitem{cabric2007cognitive}
D.~B. Cabric, ``Cognitive radios: System design perspective,'' Ph.D. dissertation, University of California, Berkeley, 2007.

\bibitem{zheng2016energy}
M.~Zheng, L.~Chen, W.~Liang, H.~Yu, and J.~Wu, ``Energy-efficiency maximization for cooperative spectrum sensing in cognitive sensor networks,'' \emph{IEEE Transactions on Green Communications and Networking}, vol.~1, no.~1, pp. 29--39, 2016.

\bibitem{wu2022energy}
Q.~Wu, B.~K. Ng, and C.-T. Lam, ``Energy-efficient cooperative spectrum sensing using machine learning algorithm,'' \emph{Sensors}, vol.~22, no.~21, p. 8230, 2022.

\bibitem{ejaz2015energy}
W.~Ejaz, G.~A. Shah, N.~u. Hasan, and H.~S. Kim, ``Energy and throughput efficient cooperative spectrum sensing in cognitive radio sensor networks,'' \emph{Transactions on Emerging Telecommunications Technologies}, vol.~26, no.~7, pp. 1019--1030, 2015.

\bibitem{deng2011energy}
R.~Deng, J.~Chen, C.~Yuen, P.~Cheng, and Y.~Sun, ``Energy-efficient cooperative spectrum sensing by optimal scheduling in sensor-aided cognitive radio networks,'' \emph{IEEE Transactions on Vehicular Technology}, vol.~61, no.~2, pp. 716--725, 2011.

\bibitem{wang2011energy}
S.~Wang, Y.~Wang, J.~P. Coon, and A.~Doufexi, ``Energy-efficient spectrum sensing and access for cognitive radio networks,'' \emph{IEEE transactions on vehicular technology}, vol.~61, no.~2, pp. 906--912, 2011.

\bibitem{ding2022deep}
X.~Ding, T.~Ni, Y.~Zou, and G.~Zhang, ``Deep learning for satellites based spectrum sensing systems: A low computational complexity perspective,'' \emph{IEEE Transactions on Vehicular Technology}, vol.~72, no.~1, pp. 1366--1371, 2022.

\bibitem{mei2023deep}
R.~Mei and Z.~Wang, ``Deep learning-based wideband spectrum sensing: A low computational complexity approach,'' \emph{IEEE Communications Letters}, 2023.

\bibitem{alam2013computationally}
S.~S. Alam, M.~O. Mughal, L.~Marcenaro, and C.~S. Regazzoni, ``Computationally efficient compressive sensing in wideband cognitive radios,'' in \emph{2013 Seventh International Conference on Next Generation Mobile Apps, Services and Technologies}.\hskip 1em plus 0.5em minus 0.4em\relax IEEE, 2013, pp. 226--231.

\bibitem{ronneberger2015u}
O.~Ronneberger, P.~Fischer, and T.~Brox, ``U-net: Convolutional networks for biomedical image segmentation,'' in \emph{Medical image computing and computer-assisted intervention--MICCAI 2015: 18th international conference, Munich, Germany, October 5-9, 2015, proceedings, part III 18}.\hskip 1em plus 0.5em minus 0.4em\relax Springer, 2015, pp. 234--241.

\bibitem{siddique2021u}
N.~Siddique, S.~Paheding, C.~P. Elkin, and V.~Devabhaktuni, ``U-net and its variants for medical image segmentation: A review of theory and applications,'' \emph{IEEE access}, vol.~9, pp. 82\,031--82\,057, 2021.

\bibitem{oktay2018attention}
O.~Oktay, J.~Schlemper, L.~L. Folgoc, M.~Lee, M.~Heinrich, K.~Misawa, K.~Mori, S.~McDonagh, N.~Y. Hammerla, B.~Kainz \emph{et~al.}, ``Attention u-net: Learning where to look for the pancreas,'' \emph{arXiv preprint arXiv:1804.03999}, 2018.

\bibitem{nguyen2024enhancing}
D.-H. Nguyen, T.-V. Nguyen, and T.~Huynh-The, ``Enhancing spectrum sensing for 5g and lte with improved u-net architecture,'' in \emph{2024 International Conference on Advanced Technologies for Communications (ATC)}.\hskip 1em plus 0.5em minus 0.4em\relax IEEE, 2024, pp. 167--172.

\bibitem{uvaydov2024stitching}
D.~Uvaydov, M.~Zhang, C.~P. Robinson, S.~D’Oro, T.~Melodia, and F.~Restuccia, ``Stitching the spectrum: Semantic spectrum segmentation with wideband signal stitching,'' in \emph{IEEE INFOCOM 2024-IEEE Conference on Computer Communications}.\hskip 1em plus 0.5em minus 0.4em\relax IEEE, 2024, pp. 2219--2228.

\bibitem{dakic2024spiking}
K.~Dakic, B.~Al~Homssi, and A.~Al-Hourani, ``Spiking-unet: Spiking neural networks for spectrum occupancy monitoring,'' in \emph{2024 IEEE Wireless Communications and Networking Conference (WCNC)}.\hskip 1em plus 0.5em minus 0.4em\relax IEEE, 2024, pp. 1--6.

\bibitem{park2022target}
J.-H. Park, D.-H. Park, and H.-N. Kim, ``Target detection using u-net for a dtv-based passive bistatic radar system,'' in \emph{2022 International Conference on Artificial Intelligence in Information and Communication (ICAIIC)}.\hskip 1em plus 0.5em minus 0.4em\relax IEEE, 2022, pp. 176--178.

\bibitem{west2021wideband}
N.~West, T.~Roy, and T.~O'Shea, ``Wideband signal localization with spectral segmentation,'' \emph{arXiv preprint arXiv:2110.00583}, 2021.

\bibitem{leemis1995reliability}
L.~M. Leemis, \emph{Reliability: probabilistic models and statistical methods}.\hskip 1em plus 0.5em minus 0.4em\relax Prentice-Hall, Inc., 1995.

\end{thebibliography}

\begin{IEEEbiography}[{\includegraphics[width=1in,height=1.25in,clip,keepaspectratio]{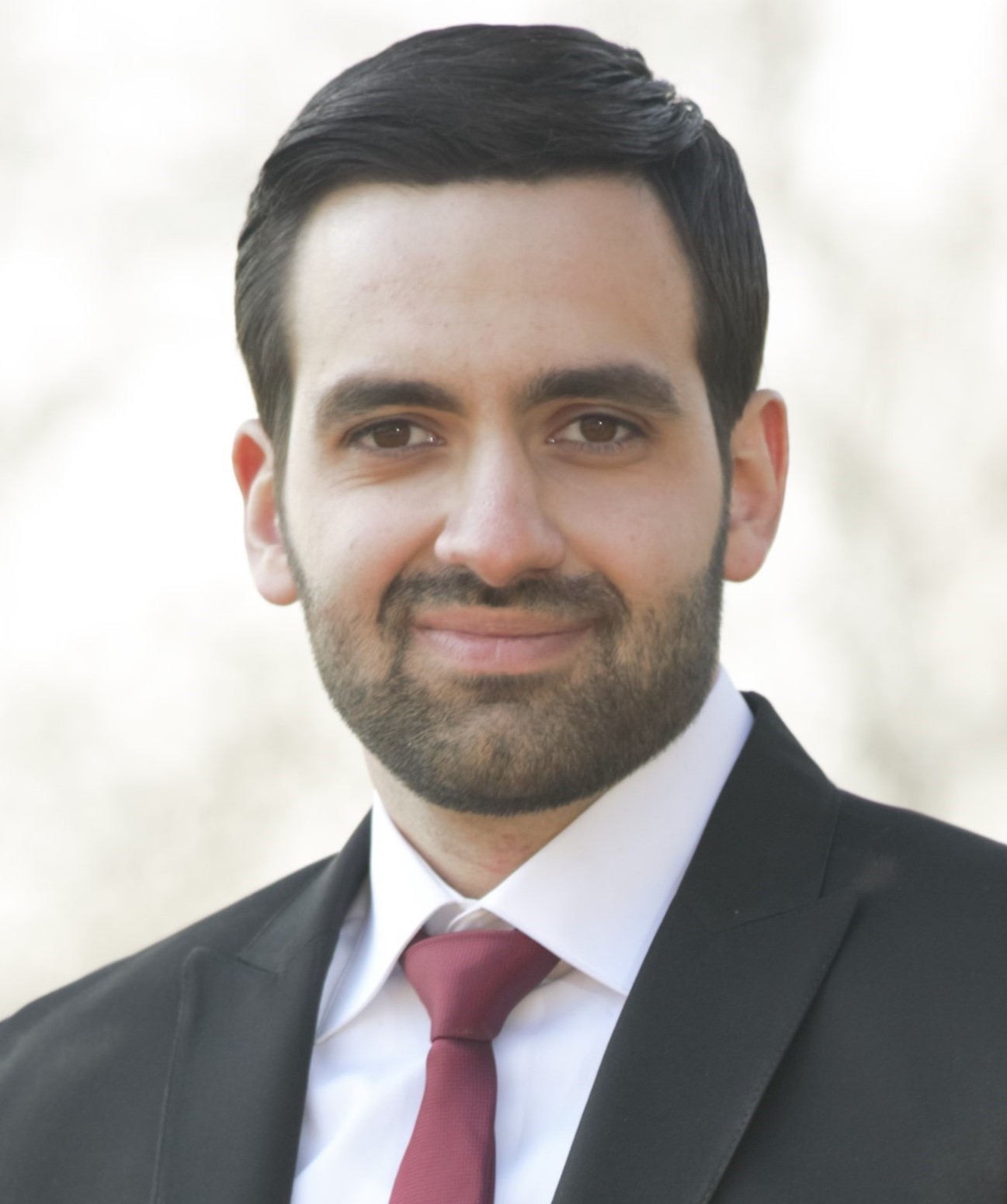}}]{Ali Rasteh }
received B.Sc. and M.Sc. degrees in electrical and computer engineering from Sharif University of Technology, Tehran, Iran, in 2015 and 2017, respectively. He is currently pursuing a Ph.D. degree in electrical and computer engineering at New York University, Brooklyn, NY, USA, where he is a member of the NYU Wireless research center. His major field of study is wireless communications.

He is currently a Research Assistant at NYU Wireless, where he works on wireless spectrum sensing, hardware development for communication systems, and AI/ML applications for 6G wireless. From 2015 to 2022, he held multiple positions at Sina Communication Systems Co., including Technical Lead and Hardware/Embedded Systems Developer, where he led development teams on advanced optical access and transport network hardware. He has also conducted part-time research with CNRS and the University of Toulouse, focusing on spiking neural networks. He has authored or co-authored several technical papers. His current research interests include 6G communication systems, efficient hardware design, and machine learning for wireless communications.

Mr. Rasteh is a member of NYU Wireless and the NYU Center for Advanced Technology in Telecommunications (CATT). He has been awarded full scholarships for his doctoral studies at New York University. He has contributed to numerous high-impact academic and industrial projects in communication technologies.
\end{IEEEbiography}

\begin{IEEEbiography}
[{\includegraphics[width=1in,height=1.25in,clip,keepaspectratio]{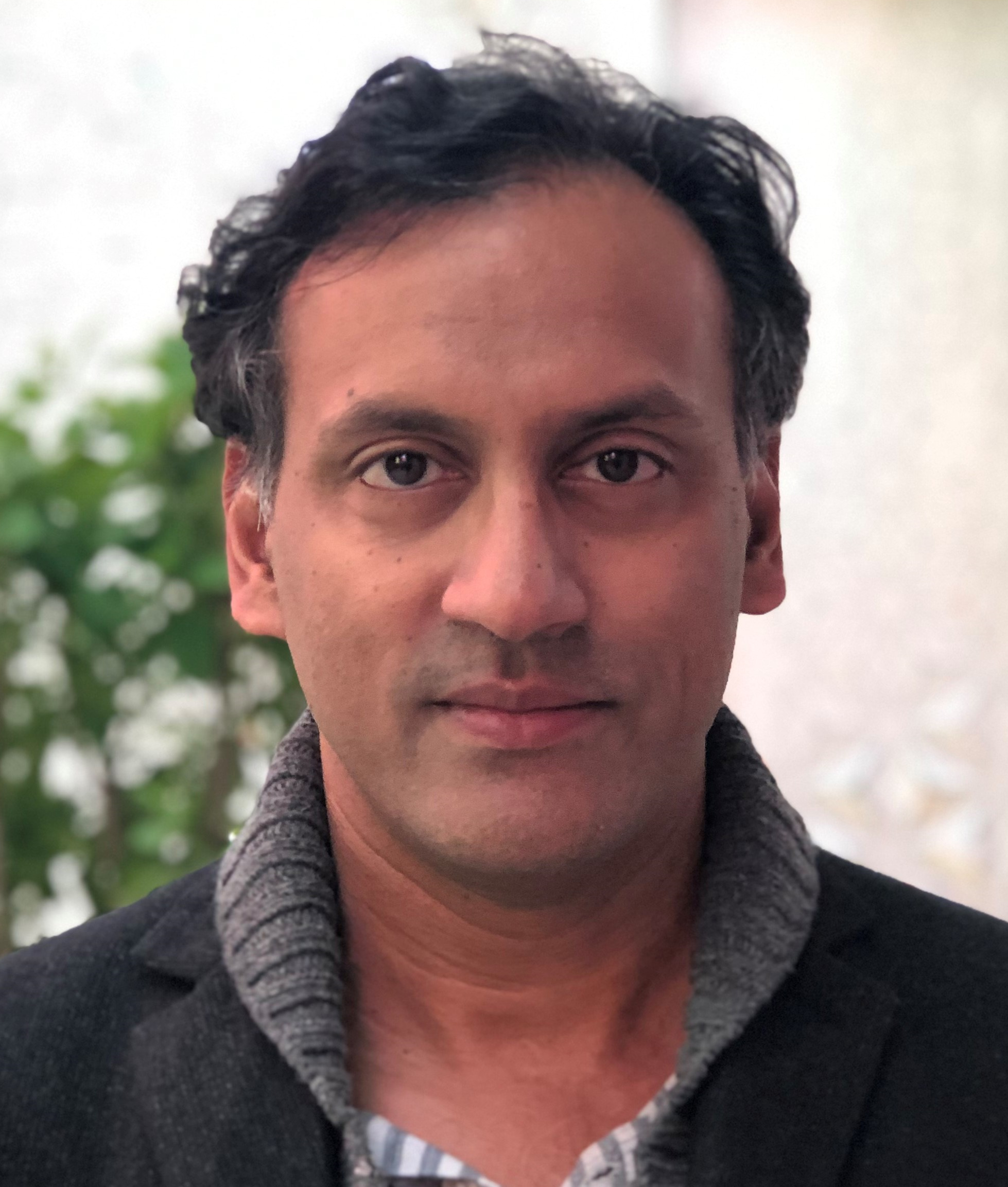}}]{Sundeep Rangan }
(Fellow, IEEE) received the
B.A.Sc. degree in electrical engineering from the
University of Waterloo, Waterloo, ON, Canada, and
the M.Sc. and Ph.D. degrees in electrical engineering
from the University of California, Berkeley, Berkeley,
CA, USA.

He has held Postdoctoral appointments
with the University of Michigan, Ann Arbor, Ann Arbor, MI, USA, and Bell Labs. In 2000, he co-founded
(with four others) Flarion Technologies, a spin-off
of Bell Labs, that developed Flash OFDM, the first
cellular OFDM data system and pre-cursor to 4G
cellular systems including LTE and WiMAX. In 2006, Flarion was acquired by
Qualcomm Technologies. He was a Senior Director of Engineering at Qualcomm
involved in OFDM infrastructure products. He joined New York University
Tandon (formerly NYU Polytechnic), Brooklyn, NY, USA, in 2010, where he is
currently a Professor of electrical and computer engineering. He is the Associate
Director of NYU WIRELESS, Brooklyn, NY, an industry-academic research
center on next-generation wireless systems.
\end{IEEEbiography}


\end{document}